\documentclass[preprint,12pt]{elsarticle}
\biboptions{sort&compress}
\usepackage{amssymb}
\usepackage{amsmath}

\journal{Materials Science in Semiconductor Processing}

\begin{document}

\begin{frontmatter}

\title{Post-diffusion cooling effects on Hall-derived active lithium donor profiles in high-purity germanium}

\author[usd]{Kunming Dong}
\author[usd]{Dongming Mei\corref{cor1}}
\ead{dongming.mei@usd.edu}
\author[usd]{Anupama Karki}
\author[usd]{Patrick Burns}
\author[usd]{Sanjay Bhattarai}

\cortext[cor1]{Corresponding author.}

\affiliation[usd]{
    organization={Department of Physics, University of South Dakota},
    addressline={414 E. Clark Street},
    city={Vermillion},
    postcode={57069},
    state={SD},
    country={USA}
}

%% Abstract
\begin{abstract}
Lithium (Li) diffusion is commonly used to form $n^{+}$ contacts in high-purity germanium (HPGe) detectors, but the final electrically active donor profile can be sensitive to the post-diffusion thermal history. Li was introduced into HPGe coupons using a lithium-in-oil suspension and diffused for $30~\mathrm{min}$ at nominal temperatures of $240$--$310~^{\circ}\mathrm{C}$. Short- and long-cooling protocols were documented by measured witness-Ge cooling histories. Sequential material removal combined with Hall-effect measurements at $77~\mathrm{K}$ was used to reconstruct difference-derived apparent Hall donor profiles. Because Hall response in a nonuniform conducting layer is mobility weighted, these profiles are operational electrically active-donor metrics rather than direct local or total-Li concentration profiles. Complementary-error-function fits were used to obtain the extrapolated apparent intercept $N_{s,\mathrm{app}}$ and the apparent profile-width parameter $D_{\mathrm{app}}$. For the coupons studied, short cooling was associated with larger $N_{s,\mathrm{app}}$ and sharper profiles, whereas long cooling was associated with lower $N_{s,\mathrm{app}}$ and broader low-concentration tails. Fit-derived concentration-threshold depths likewise extended farther into the Ge bulk for the long-cooling coupons. These results show that the complete post-diffusion thermal history should be considered when parameterizing Hall-active Li-diffused $n^{+}$ contacts for HPGe detector fabrication.
\end{abstract}

%%Graphical abstract
\begin{graphicalabstract}
\includegraphics[width=\textwidth]{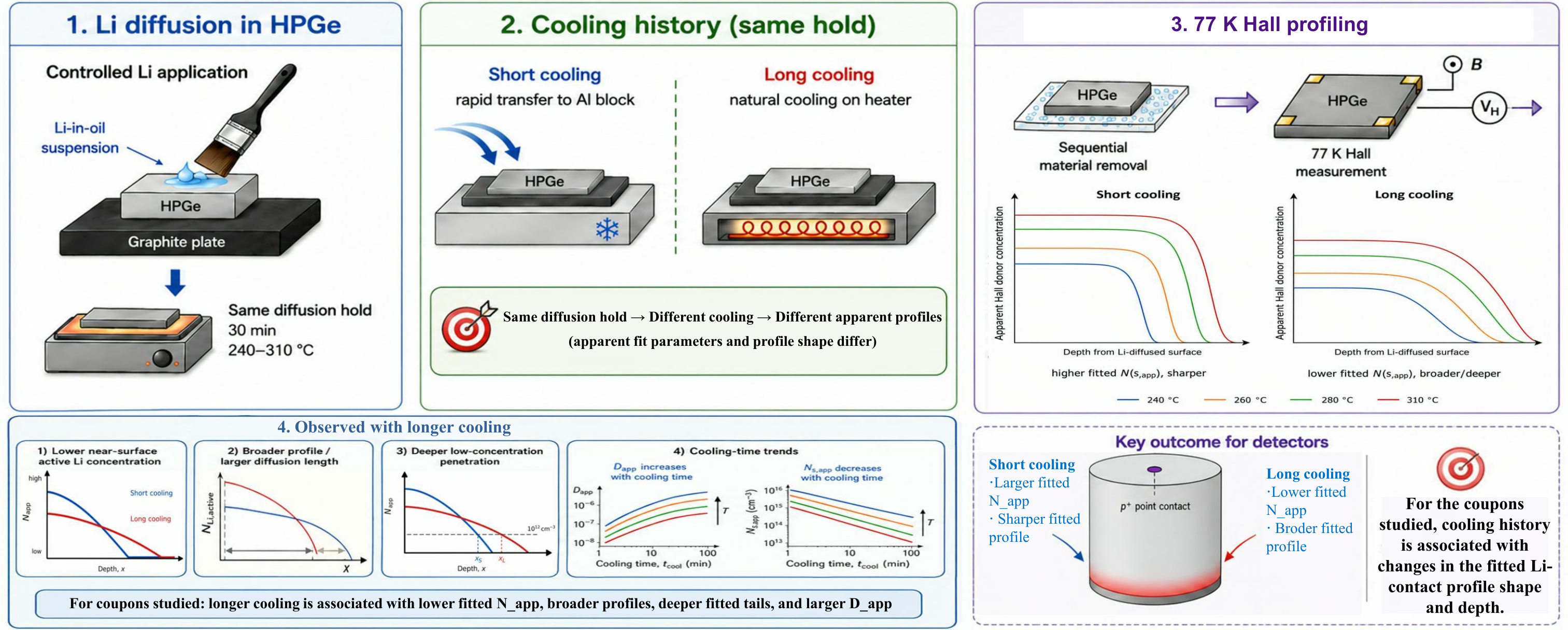}
\end{graphicalabstract}

%%Research highlights

\begin{highlights}
\item Measured cooling histories were compared with Hall-derived active Li profiles.
\item Sequential Hall measurements mapped mobility-weighted apparent donor profiles.
\item Short cooling was associated with a larger extrapolated $N_{s,\mathrm{app}}$.
\item Long cooling was associated with a broader low-concentration donor tail.
\item Apparent fit parameters are distinguished from intrinsic Li transport quantities.
\end{highlights}

%% Keywords

\end{frontmatter}

%% Add \usepackage{lineno} before \begin{document} and uncomment 
%% following line to enable line numbers
%% \linenumbers

%% main text
%%

%% Use \section commands to start a section
\section{Introduction}

High-purity germanium (HPGe) detectors are widely used in gamma-ray spectroscopy and rare-event searches because of their excellent energy resolution and low net impurity concentration \cite{knoll2010radiation,gilmore2008practical,agostini2020gerda,arnquist2023majorana}. In p-type HPGe detectors, lithium diffusion is commonly used to form the outer $n^{+}$ contact \cite{aguayo2013signals,pitale2021planar,ghosh2025fabrication}. This Li-diffused region provides a low-resistance $n^{+}$ electrode and a hole-blocking surface, but it also introduces a near-surface layer with reduced or incomplete charge collection \cite{aguayo2013signals,jiang2016deadlayer,ma2017inactive,dai2023modeling}. The thickness and gradient of this layer affect the active detector volume, the surface-event response, and the accuracy of detector simulations \cite{jiang2016deadlayer,ma2017inactive,dai2023modeling}. Therefore, controlling the Li profile is an important processing problem for HPGe detector fabrication.

Lithium diffusion in germanium (Ge) has been studied for several decades \cite{fuller1953diffusion,fuller1954mobility,pratt1966diffusion}. Early work treated Li as a highly mobile interstitial donor and described its temperature-dependent diffusivity using Arrhenius-type relations \cite{fuller1953diffusion,fuller1954mobility,pratt1966diffusion}. Fuller and Ditzenberger measured the diffusion constant using a junction-depth method and reported, for Ge, \cite{fuller1953diffusion}
\begin{equation}
D_{\mathrm{FD}}(T)=13\times 10^{-4}
\exp\left(-\frac{10700}{RT}\right)
~\mathrm{cm^{2}\,s^{-1}} .
\label{eq:fd_ge}
\end{equation}
Fuller and Severiens later used an ion-drift method and obtained, for Ge, \cite{fuller1954mobility}
\begin{equation}
D_{\mathrm{FS}}(T)=25\times 10^{-4}
\exp\left(-\frac{11800}{RT}\right)
~\mathrm{cm^{2}\,s^{-1}} .
\label{eq:fs_ge}
\end{equation}
Here, $T$ is the absolute temperature and $R$ is the gas constant in units consistent with the activation energies. These studies established the basic thermally activated behavior of Li transport in Ge, although the measured parameters depended on the experimental method and the temperature range \cite{fuller1953diffusion,fuller1954mobility,pratt1966diffusion}. Compared with Li diffusion in silicon (Si), quantitative measurements in Ge are more difficult, especially for solubility-related behavior, because Li in Ge is highly mobile and the final active profile can be affected by redistribution during the thermal history \cite{fuller1953diffusion,pratt1966diffusion,pell1957solubility,morin1957precipitation,carter1960kinetics,sangster1997geli}.

Pratt and Friedman introduced a depth-profiling method that is especially relevant to the present work \cite{pratt1966diffusion}. In their experiment, Li was diffused into Ge, thin layers were successively removed by lapping, and the conductance of the remaining diffused layer was measured after each removal step \cite{pratt1966diffusion}. The resulting impurity profile was fitted with the complementary-error-function ($\mathrm{erfc}$) solution for diffusion from an effectively constant surface source \cite{pratt1966diffusion,crank1975mathematics,shewmon1989diffusion,mehrer2007diffusion},
\begin{equation}
N(x,t)=N_{0}\mathrm{erfc}
\left(
\frac{x}{2\sqrt{Dt}}
\right) .
\label{eq:pf_profile}
\end{equation}
Using this method, they obtained the following Arrhenius relation for Li diffusion in Ge: \cite{pratt1966diffusion}
\begin{equation}
D_{\mathrm{PF}}(T)=9.10\times 10^{-3}
\exp\left(-\frac{13100}{RT}\right)
~\mathrm{cm^{2}\,s^{-1}} .
\label{eq:pf_ge}
\end{equation}
Their work showed that repeated layer removal can provide a diffusion profile with many experimental depth points, rather than relying only on a single junction-depth value \cite{pratt1966diffusion}. It also showed that Li profiles in Ge can be reasonably described by an erfc-type diffusion solution under controlled thermal conditions \cite{pratt1966diffusion,crank1975mathematics,shewmon1989diffusion,mehrer2007diffusion}.

For detector processing, however, the relevant quantity is not only the intrinsic isothermal diffusion coefficient \cite{fuller1953diffusion,fuller1954mobility,pratt1966diffusion}. A real HPGe fabrication cycle includes heating, the nominal diffusion hold, transfer from the heat source, and cooling to room temperature. During this non-isothermal part of the process, Li may continue to redistribute \cite{fuller1953diffusion,fuller1954mobility,pratt1966diffusion}. In addition, the electrically active donor concentration may change because of temperature-dependent solubility, precipitation, compensation, or the formation of electrically inactive Li-related complexes \cite{pell1957solubility,morin1957precipitation,reiss1958effect,carter1960kinetics,sangster1997geli}. These effects are important because detector electrostatics are governed by the electrically active donor profile, not necessarily by the total Li concentration introduced into the crystal \cite{sze2007physics,knoll2010radiation,aguayo2013signals,dai2023modeling}.

In this work, we study Li diffusion in HPGe coupons under controlled conditions. Diffusion temperatures were selected within a relatively narrow range from $240~^{\circ}\mathrm{C}$ to $310~^{\circ}\mathrm{C}$. This range is relevant to practical Li contact formation in HPGe detector fabrication, where sufficient Li penetration is required without unnecessary high-temperature exposure \cite{pratt1966diffusion,aguayo2013signals,pitale2021planar,ghosh2025fabrication}. Operation substantially above this range can increase the risk of unwanted impurity redistribution or contamination-related effects in high-purity Ge \cite{yang2013annealing,morin1957precipitation,carter1960kinetics,sze2007physics}. The upper end of the range was therefore kept close to the temperatures normally used for Li diffusion, while allowing the temperature dependence of the final active profile to be evaluated \cite{pratt1966diffusion,pitale2021planar,ghosh2025fabrication}.

Li was introduced from an in-house lithium-in-oil suspension and diffused with a fixed nominal hold time. Two different cooling protocols were used to examine the effect of post-diffusion thermal history. After diffusion, apparent active Li donor profiles were obtained by sequential material removal from the Li-diffused side and Hall-effect measurements at $77~\mathrm{K}$ \cite{pratt1966diffusion,yang2013annealing,mei2016neutral,mei2017neutral,raut2020characterization}. The term ``apparent'' is used because Hall measurements on samples with nonuniform donor profiles do not directly measure the local total Li concentration at a single newly exposed depth. Instead, the Hall-depth-profiling results are interpreted as Hall-derived electrically active donor profiles obtained from the sequential material-removal procedure.

The Hall measurements were performed at $77~\mathrm{K}$ because HPGe detectors are commonly operated near liquid-nitrogen temperature \cite{knoll2010radiation,gilmore2008practical,lutz1999semiconductor}. This temperature is also advantageous for Hall measurements in high-purity Ge: at room temperature, thermally generated intrinsic carriers can dominate the measured carrier concentration because of the small band gap of Ge, while cooling to $77~\mathrm{K}$ strongly suppresses this contribution \cite{sze2007physics,shur1990physics,iwan1982highpurity,mei2016neutral,mei2017neutral,raut2020characterization}. The resulting measurement is therefore more directly related to the electrically active donor profile relevant to cryogenic detector operation \cite{yang2013annealing,mei2016neutral,mei2017neutral,raut2020characterization}.

The measured profiles are used to extract process-dependent apparent Li profile parameters for the different cooling protocols. In this paper, the extrapolated apparent erfc intercept and profile-width parameter are denoted as $N_{s,\mathrm{app}}$ and $D_{\mathrm{app}}$, respectively. The former is a fit extrapolation to $x=0$, not a direct measurement of the surface donor concentration, and the latter is not an intrinsic isothermal diffusivity. These parameters are compared with classical Arrhenius descriptions of Li diffusion in Ge to evaluate how the complete processing history is reflected in the final Hall-derived profile \cite{fuller1953diffusion,fuller1954mobility,pratt1966diffusion}. The measured short- and long-cooling endpoints are the primary experimental comparisons; any interpolation between those endpoints is used only as a descriptive visualization and is not treated as an independently validated transport law.

This study is part of a broader HPGe detector-development program in which our laboratory has established an integrated platform for Ge material preparation, crystal growth, detector machining, contact formation, passivation, and cryogenic detector characterization \cite{wang2015crystal,bhattarai2024crystal,wei2018contact,meng2019fabrication,panth2020cryogenic,raut2020characterization}. The present paper focuses on the materials-processing problem that underlies such devices: controlling the Li-diffused $n^{+}$ contact in HPGe.

The goal of this work is to connect classical Li diffusion concepts with practical HPGe detector fabrication \cite{fuller1953diffusion,fuller1954mobility,pratt1966diffusion,aguayo2013signals}. The analysis is deliberately separated into three levels: (i) the experimentally observed Hall-derived profile differences between the short- and long-cooling coupons, (ii) an erfc parameterization of those profiles using the apparent quantities $N_{s,\mathrm{app}}$ and $D_{\mathrm{app}}$, and (iii) discussion of physical mechanisms that may contribute to the observed differences. This distinction avoids assigning a unique microscopic mechanism to the Hall data alone. The resulting process-level description is relevant to $n^{+}$ contact engineering, reduced-charge-collection layers, active-volume estimation, and future fabrication of large-volume HPGe detectors \cite{aguayo2013signals,jiang2016deadlayer,ma2017inactive,dai2023modeling,dandrea2021germanium,abgrall2021legend1000}.

\section{Experimental Methods}

\subsection{HPGe samples and surface preparation}

\begin{figure}[htp!]
    \centering
    \includegraphics[width=0.72\linewidth]{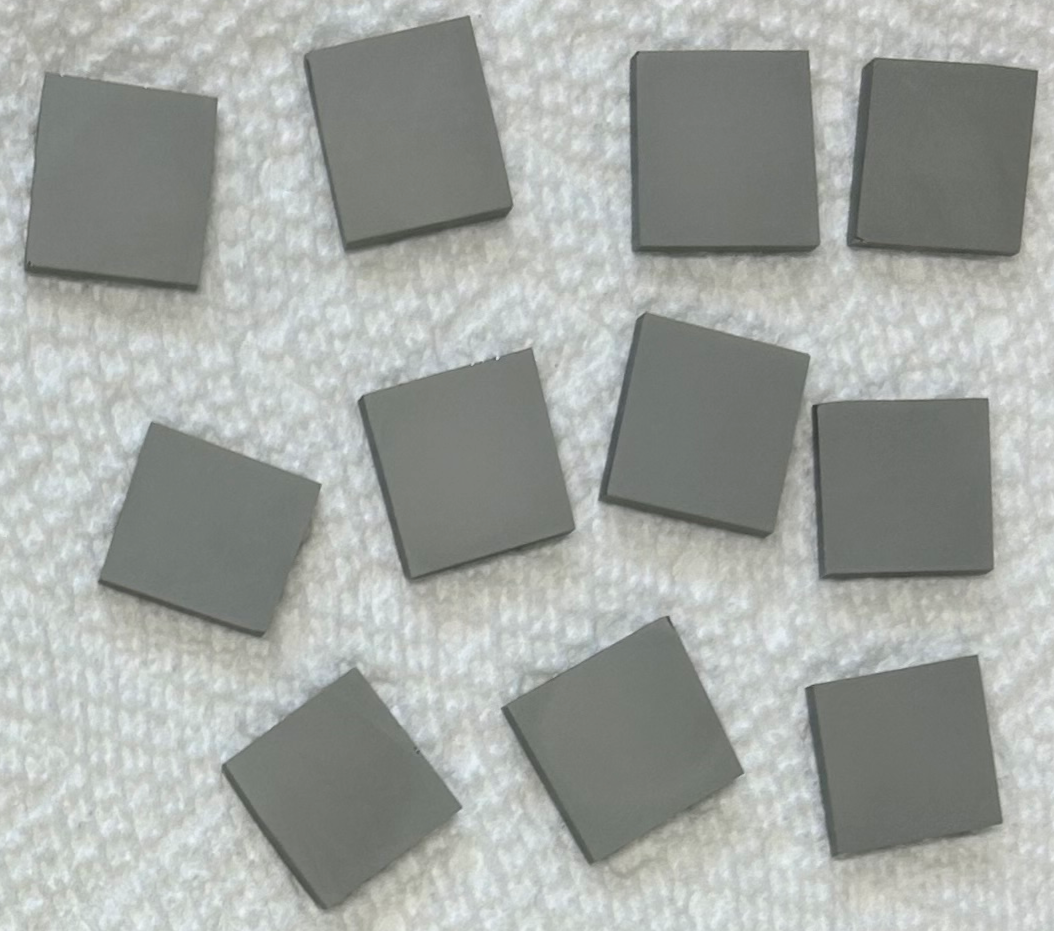}
    \caption{Representative HPGe coupon samples prepared from a wafer cut from the lower side of the HPGe crystal grown at the University of South Dakota. These coupons were used for Li application, diffusion, and subsequent apparent active donor profile measurements.}
    \label{fig:hpge_samples}
\end{figure}

The HPGe coupon samples used in this study were prepared from a p-type HPGe single crystal grown at the University of South Dakota in December 2017 \cite{wang2015crystal,bhattarai2024crystal,raut2020characterization}. The measured net acceptor concentration in the crystal region used for the coupons was below $5\times10^{10}~\mathrm{cm^{-3}}$, consistent with detector-grade HPGe. A wafer with a thickness of approximately $1.8~\mathrm{mm}$ was cut from the lower side of the crystal using a wire saw. The wafer was then divided into square coupons with lateral dimensions of approximately $1.5~\mathrm{cm}\times1.5~\mathrm{cm}$. Representative prepared coupons are shown in Fig.~\ref{fig:hpge_samples}.

Before Li application, both major surfaces of each coupon were mechanically polished following the same surface-preparation procedure used for planar HPGe detector fabrication in our laboratory \cite{wei2018contact,meng2019fabrication,panth2020cryogenic,raut2020characterization}. The surfaces were first polished with $17.5~\mu\mathrm{m}$ silicon carbide (SiC) abrasive and then with $9.5~\mu\mathrm{m}$ aluminum oxide ($\mathrm{Al_2O_3}$) abrasive. Polishing was continued until no optically visible scratches remained on the major surfaces. If scratches were observed after chemical etching, the final $9.5~\mu\mathrm{m}$ $\mathrm{Al_2O_3}$ polishing step was repeated.

After mechanical polishing, the coupons were chemically etched in a $4{:}1$ mixture of nitric acid and hydrofluoric acid, $\mathrm{HNO_3{:}HF}$, for approximately $3~\mathrm{min}$ \cite{wei2018contact,meng2019fabrication,panth2020cryogenic,raut2020characterization}. During etching, the coupons were gently agitated to promote uniform removal of the mechanically damaged surface layer. The etched coupons were then thoroughly rinsed with deionized water and dried with nitrogen gas. Immediately before transfer for Li application, each coupon received an additional short etch in the same $4{:}1$ $\mathrm{HNO_3{:}HF}$ solution for approximately $30~\mathrm{s}$, followed by deionized-water rinsing and nitrogen drying \cite{wei2018contact,meng2019fabrication,panth2020cryogenic,raut2020characterization}.

The prepared coupons were handled with rubber-coated tweezers to minimize surface scratching and metallic contamination. After the final drying step, the coupons were placed on a high-purity graphite plate, evacuated, and transferred into a glovebox continuously circulated with ultra-high-purity argon (Ar). The coupons were kept in the Ar environment until Li application and diffusion.

\subsection{Lithium application and diffusion protocols}

Lithium was applied to one major surface of each HPGe coupon using an in-house lithium-in-oil suspension, prepared following the procedure described in our previous work~\cite{dong2026hybrid}. The suspension consisted of Li particles dispersed in mineral oil. Immediately before application, the suspension was gently stirred with a small brush to redistribute Li particles that had separated from the oil during storage. A fine brush was then used to apply a thin and visually uniform layer of the Li suspension onto the upper surface of each coupon. Care was taken to cover the surface intended for Li diffusion while avoiding unnecessary spreading onto the sidewalls. The application criterion was visual uniformity rather than an independently quantified Li areal dose for each coupon. Consequently, condition-to-condition source-loading variation cannot be excluded and is treated as one of the experimental limitations when interpreting differences in the extrapolated $N_{s,\mathrm{app}}$ and fit-derived apparent dose.

\begin{figure}[htp!]
    \centering
    \includegraphics[width=0.78\linewidth]{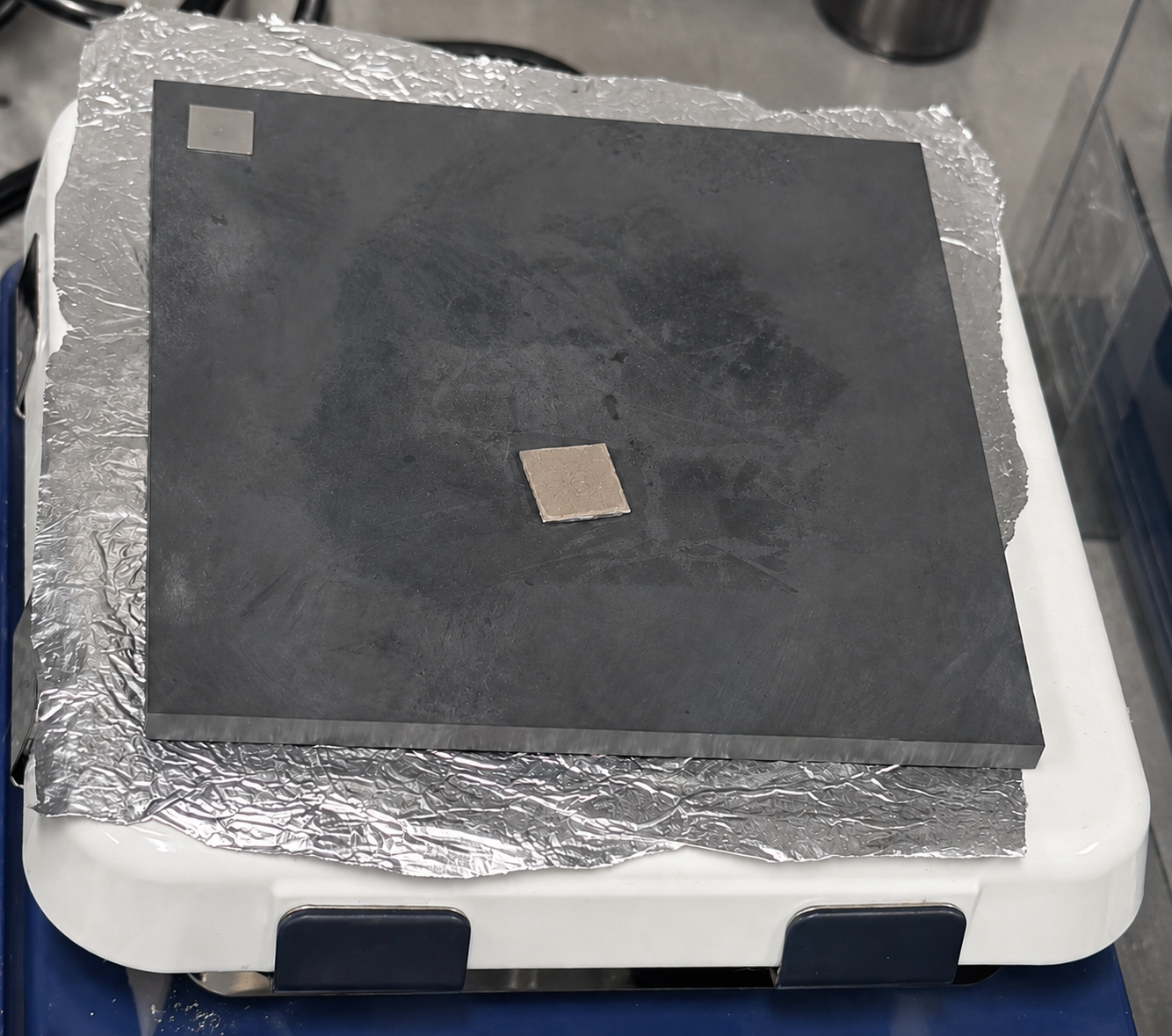}
    \caption{Li diffusion setup showing an HPGe coupon placed on a high-purity graphite plate. A separate Ge piece located near the corner of the graphite plate was used as a witness sample for temperature monitoring during diffusion.}
    \label{fig:li_diffusion_setup}
\end{figure}

The diffusion was carried out on a high-purity graphite plate, as shown in Fig.~\ref{fig:li_diffusion_setup}. A separate Ge piece was placed near one corner of the graphite plate and used as a temperature-monitoring witness sample. This arrangement allowed the process temperature to be monitored on Ge under the same nominal thermal environment as the Li-coated coupon, without disturbing the coated diffusion surface. Before each diffusion run, the heater was preheated for approximately $5~\mathrm{min}$ to reduce the initial thermal transient. The graphite plate carrying the Li-coated coupon and the Ge witness sample was then transferred onto the heater.

Each sample was diffused at the selected set-point temperature for a nominal hold time of $30~\mathrm{min}$. Before processing the eight detector-grade coupons, the heater settings had been established through repeated practice runs using non-detector-grade Ge samples. During each reported hold, the temperature was monitored every $2~\mathrm{min}$ using a Hanna HI935005 thermometer equipped with a HI766B surface probe. The probe was placed on the Ge witness sample rather than directly on the Li-coated coupon. The measured witness-sample temperature remained stable during each diffusion hold, indicating temporally stable heating of the coupon-scale setup; with one coupon per condition, this observation is not a test of run-to-run process reproducibility. Because a sample-specific time-averaged coupon temperature was not independently retained, the temperatures reported for the eight coupons are nominal process set points rather than measured mean coupon temperatures.

At the end of the $30~\mathrm{min}$ diffusion hold, the heater was switched off and the sample was cooled according to the cooling protocol assigned to that run. The cooling step was treated as part of the overall thermal history because Li can continue to redistribute while the sample remains at elevated temperature~\cite{fuller1953diffusion,fuller1954mobility,pratt1966diffusion,pell1957solubility,morin1957precipitation,carter1960kinetics}. The specific cooling protocols are described in the following subsection.

\subsection{Cooling protocols}

Two cooling protocols were used to evaluate the effect of post-diffusion thermal history on the final electrically active Li profile~\cite{fuller1953diffusion,fuller1954mobility,pratt1966diffusion,pell1957solubility,morin1957precipitation,carter1960kinetics}. The cooling-time coordinate, $t_{\mathrm{cool}}$, was defined with $t_{\mathrm{cool}}=0$ at the end of the nominal $30~\mathrm{min}$ diffusion hold. In the long-cooling protocol, the heater was switched off and the graphite plate carrying the HPGe coupon was left on the heater to cool naturally. In the short-cooling protocol, the graphite plate was rapidly transferred from the heater to a large aluminum (Al) cold stage immediately after the diffusion hold. The Al cold stage provided a substantially larger thermal mass and reduced the time during which the sample remained at elevated temperature. For each cooling protocol, the scalar endpoint cooling time, $t_{\mathrm{cool},30}$, was defined as the elapsed time required for the measured witness Ge temperature to reach approximately $30~^{\circ}\mathrm{C}$.

\begin{figure}[htp!]
    \centering
    \includegraphics[width=\textwidth]{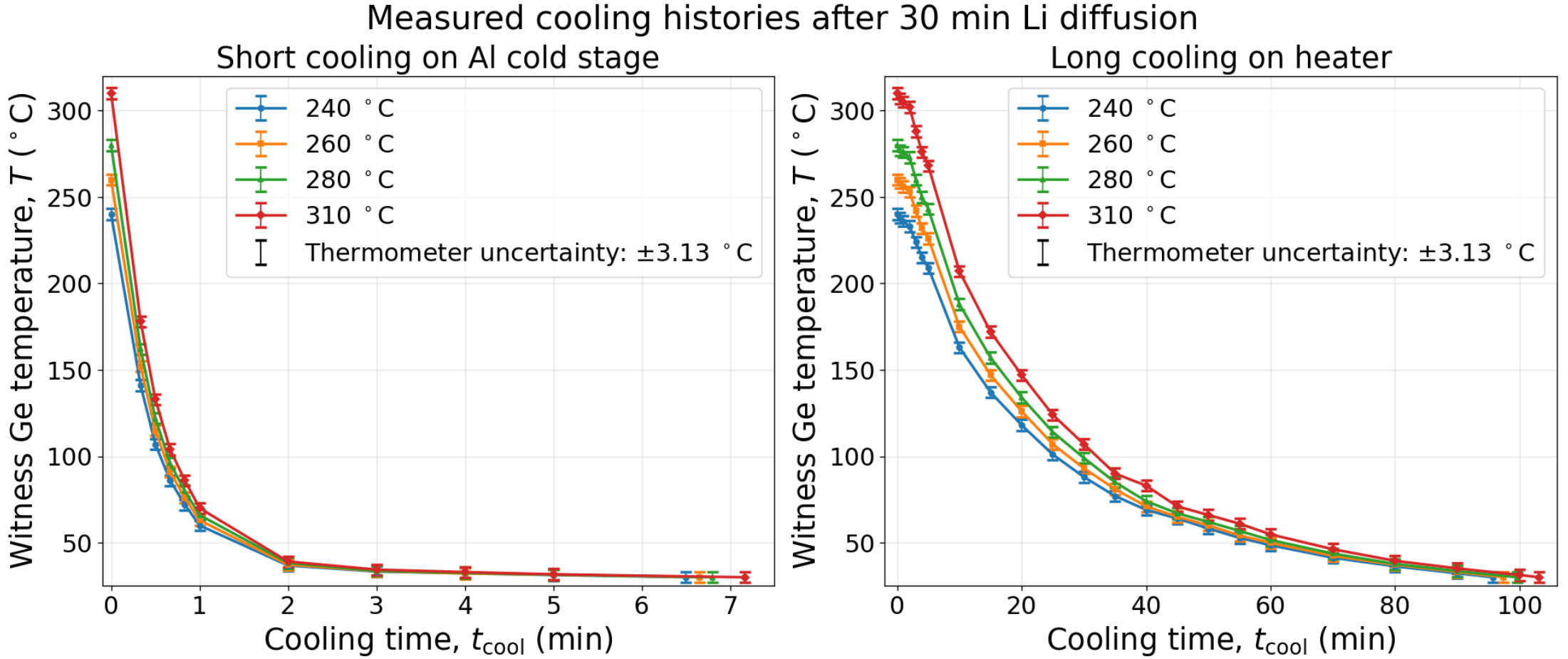}
    \caption{Measured cooling histories after the $30~\mathrm{min}$ Li diffusion hold. The left panel shows the short-cooling protocol, in which the graphite plate carrying the HPGe coupon was transferred to an Al cold stage. The right panel shows the long-cooling protocol, in which the graphite plate remained on the heater after the heater was switched off. The time origin corresponds to the end of the nominal diffusion hold. Vertical error bars represent thermometer/probe measurement uncertainty of $\pm 3.13~^{\circ}\mathrm{C}$, calculated as $\sigma_T=\sqrt{(2.7)^2+(1.5)^2+(0.5)^2}~^{\circ}\mathrm{C}$. This uncertainty includes the thermometer meter contribution, surface-probe uncertainty, and temperature-readout resolution, but does not include possible systematic differences between the witness Ge sample and the Li-coated coupon or unresolved thermal transients during rapid transfer.}
    \label{fig:cooling_history}
\end{figure}

For both protocols, the temperature was monitored on the Ge witness sample placed on the graphite plate, using the same Hanna HI935005 thermometer and HI766B surface probe described above. The measured cooling histories are shown in Fig.~\ref{fig:cooling_history}. The short-cooling protocol reduced the witness Ge temperature from the diffusion temperature to approximately $60$--$70~^{\circ}\mathrm{C}$ within $1~\mathrm{min}$, to approximately $37$--$39~^{\circ}\mathrm{C}$ within $2~\mathrm{min}$, and to approximately $30~^{\circ}\mathrm{C}$ within about $6.5$--$7.2~\mathrm{min}$. In contrast, the long-cooling protocol kept the witness Ge sample at much higher temperature for a longer time. After $20~\mathrm{min}$ of long cooling, the measured witness Ge temperatures were approximately $118$, $126$, $134$, and $147~^{\circ}\mathrm{C}$ for the $240$, $260$, $280$, and $310~^{\circ}\mathrm{C}$ diffusion runs, respectively, and the time required to reach approximately $30~^{\circ}\mathrm{C}$ was about $96$--$103~\mathrm{min}$.

The cooling step was treated as an intentional process variable rather than as an uncontrolled experimental detail. Although the nominal Li diffusion hold was identical for samples processed at the same set-point temperature, Li can continue to diffuse during cooling while the sample remains warm. Therefore, the physically relevant record is the complete measured temperature trajectory $T(t)$, not only the elapsed time required to reach $30~^{\circ}\mathrm{C}$~\cite{fuller1953diffusion,pratt1966diffusion,crank1975mathematics,shewmon1989diffusion,mehrer2007diffusion}. For a chosen literature diffusivity $D_{\mathrm{lit}}(T)$, a diffusion-weighted cooling exposure can be written as
\begin{equation}
\Theta_{\mathrm{cool}}^{\mathrm{lit}}
=
\int_{\mathrm{cooling}} D_{\mathrm{lit}}[T(t)]\,dt,
\label{eq:cooling_thermal_exposure}
\end{equation}
and the corresponding classical total thermal exposure as
\begin{equation}
\Theta_{\mathrm{tot}}^{\mathrm{lit}}
=
D_{\mathrm{lit}}(T_{\mathrm{hold}})t_{\mathrm{hold}}
+
\Theta_{\mathrm{cool}}^{\mathrm{lit}}.
\label{eq:total_thermal_exposure}
\end{equation}
These expressions clarify why a long period near room temperature contributes far less to classical diffusion than the earlier, hotter part of the cooling curve. In the present study, however, the temperature was measured on a witness Ge piece rather than directly on the Li-coated coupon, and unresolved rapid-transfer transients are not quantified. We therefore do not convert Eqs.~\eqref{eq:cooling_thermal_exposure} and~\eqref{eq:total_thermal_exposure} into an intrinsic diffusion length or use them to correct $D_{\mathrm{app}}$. Accordingly, Eqs.~\eqref{eq:cooling_thermal_exposure} and~\eqref{eq:total_thermal_exposure} are used as a physical interpretation framework rather than as a numerically evaluated correction to the measured Hall-derived profiles. Instead, Fig.~\ref{fig:cooling_history} documents the full measured thermal trajectories, while the scalar endpoint time $t_{\mathrm{cool},30}$ is used later only as a descriptive process coordinate. It should not be interpreted as an equivalent isothermal diffusion time.

\subsection{Sequential material removal and $77~\mathrm{K}$ Hall profiling}

After completion of the assigned cooling protocol, the diffused coupon was immersed in methanol to remove residual Li and mineral oil from the surface. The sample was kept in methanol until visible gas evolution ceased, indicating that the remaining reactive Li on the surface had been consumed. The coupon was then rinsed with isopropyl alcohol and dried with nitrogen gas.

The electrically active Li donor profile was estimated by repeated Hall-effect measurements combined with sequential material removal from the Li-diffused side~\cite{fuller1954mobility,sze2007physics,shur1990physics}. Because the Li-diffused coupons had strongly nonuniform donor profiles, an individual Hall measurement after a removal step was not interpreted as the local carrier concentration at the newly exposed surface. Instead, adjacent measurements were differenced to obtain a difference-derived apparent Hall donor concentration associated with the material removed between the two measurements.

Hall measurements were performed with an Ecopia HMS-3000 Hall-effect measurement system. The permanent-magnet module used for these measurements was marked $0.58~\mathrm{T}$, and the coupons were mounted on an Ecopia SPCB-01 sample board using its four spring-clip contacts. Liquid nitrogen was added through the system inlet to establish the nominal $77~\mathrm{K}$ measurement condition. The liquid-nitrogen inventory was sufficient to maintain this condition throughout an individual measurement sequence and could be replenished through the inlet when necessary.

Let $i=0$ denote the as-diffused state after surface cleaning, and let $i=1,2,\ldots$ denote subsequent polish--etch--measure steps. At each step, the coupon was measured four times with a digital caliper after changing the measurement direction, and the four readings were averaged to obtain the remaining thickness, $h_i$. Agreement among the four readings was used as an operational check of polishing uniformity. The caliper resolution was $0.01~\mathrm{mm}$; the resolution contribution used below was conservatively not reduced by the four-reading average. The cumulative depth removed from the Li-diffused surface was denoted by $x_i$, with $x_0=0$. For the material-removal interval between steps $i-1$ and $i$, the measured total thickness decrease was
\begin{equation}
    \Delta h_i = h_{i-1}-h_i .
\end{equation}
Since the chemical etch removed Ge from both major surfaces, only the Li-side part of the etch contributes to the Li-profile depth. For the standard $10~\mathrm{s}$ etch, the calibrated removal from each major surface was approximately
$e_{\mathrm{etch}}=1.6~\mu\mathrm{m}$. The Li-side depth increment was therefore calculated as
\begin{equation}
    \Delta x_i = \Delta h_i - e_{\mathrm{etch}},
\end{equation}
where the subtracted term accounts for the backside material removed by the same etch step. The cumulative Li-side depth was then
\begin{equation}
    x_i = \sum_{j=1}^{i} \Delta x_j .
\end{equation}

For each Hall measurement, the Hall system directly reported a signed bulk Hall carrier concentration, $n_{H,i}$, for the remaining coupon after removal to depth $x_i$. Because the Li-diffused coupons had nonuniform donor profiles, this value was not interpreted as the local active donor concentration at the newly exposed surface. Instead, adjacent Hall measurements were used to construct a difference-derived apparent quantity for the material removed between two successive measurements.

The apparent active Li donor concentration assigned to the removed slice between steps $i-1$ and $i$ was calculated as
\begin{equation}
    N_{\mathrm{Li,active},i}^{\mathrm{app}}
    =
    \frac{
    \left| n_{H,i-1} \right| h_{i-1}
    -
    \left| n_{H,i} \right| h_i
    }{
    x_i-x_{i-1}
    } ,
    \label{eq:hall_adjacent_difference}
\end{equation}
where $h_i$ is the remaining coupon thickness at step $i$. The signed values were retained in the source workbooks. All 166 Hall measurements across the eight profiles had the same negative carrier-concentration sign, corresponding to $n$-type conduction under the Hall-system sign convention; therefore, taking the absolute value did not conceal an observed $n$- to $p$-type transition.

This reconstructed concentration was assigned to the midpoint depth of the removed slice,
\begin{equation}
    x_{i-\frac{1}{2}}
    =
    \frac{x_{i-1}+x_i}{2}.
    \label{eq:hall_midpoint_depth}
\end{equation}
Thus, each plotted profile point is assigned to a finite removed-depth interval, but it is not necessarily the arithmetic average of the local donor concentration in that slice. For parallel $n$-type conducting layers with sheet carrier densities $S_j$ and mobilities $\mu_j$, a low-field single-carrier Hall analysis gives the effective Hall sheet density
\begin{equation}
S_H
=
\frac{\left(\sum_j S_j\mu_j\right)^2}
{\sum_j S_j\mu_j^2}.
\label{eq:hall_mobility_weighting}
\end{equation}
Only when one carrier type dominates and the Hall factors and mobilities are approximately uniform does $S_H$ reduce to the integrated carrier density $\sum_j S_j$~\cite{schroder2006semiconductor,basol2024dhem}. Equation~\eqref{eq:hall_adjacent_difference} should therefore be read as a difference of effective Hall sheet densities divided by the removed interval thickness. It is an operational, mobility-weighted profile metric rather than a direct local donor or total-Li concentration.

The recorded Hall mobilities span $9.34\times10^{3}$--$3.58\times10^{4}~\mathrm{cm^{2}\,V^{-1}\,s^{-1}}$ over the full data set, and the within-profile maximum-to-minimum ratios range from 1.8 to 3.3. The uniform-mobility condition is therefore not satisfied exactly. As an illustrative two-layer sensitivity check, if two $n$-type layers have equal sheet density and mobility ratio $r=\mu_2/\mu_1$, Eq.~\eqref{eq:hall_mobility_weighting} gives
\begin{equation}
\frac{S_H}{S_1+S_2}
=
\frac{(1+r)^2}{2(1+r^2)}.
\label{eq:two_layer_sensitivity}
\end{equation}
Using $r=1.8$--$3.3$ as an illustrative contrast gives $S_H/(S_1+S_2)\simeq0.93$--$0.78$, showing that mobility weighting can change the inferred sheet density by roughly $7$--$22\%$ even when both layers remain $n$-type. This calculation is a sensitivity estimate, not a correction applied to the reported profiles, because the local mobility distribution is not independently known.

After each Hall measurement, material was removed from the Li-diffused side by polishing with $9.5~\mu\mathrm{m}$ $\mathrm{Al_2O_3}$ abrasive. The sample was then chemically cleaned by etching for approximately $10~\mathrm{s}$ in the same $4{:}1$ $\mathrm{HNO_3{:}HF}$ solution used for the initial surface preparation, followed by deionized-water rinsing and nitrogen drying. A new Hall measurement at $77~\mathrm{K}$ and a new thickness measurement were then performed.

This polish--etch--measure sequence was repeated for each coupon until the Li-diffused region was sufficiently sampled for profile fitting. The ``Plot Data'' worksheets used by the profile-fitting notebooks had already omitted the first reconstructed surface slice, which is the slice most sensitive to residual surface Li, surface-cleaning effects, and contact-geometry artifacts. Consequently, no additional raw profile points were removed in the notebook analysis. After workbook preprocessing and exclusion of nonfinite or nonpositive values, the individual erfc fits contained 12--31 valid difference-derived profile points. Because the first surface slice is not included, $N_{s,\mathrm{app}}$ is necessarily an extrapolated erfc intercept at $x=0$ rather than a directly measured surface concentration. We therefore retain the notation ``extrapolated apparent intercept'' or $N_{s,\mathrm{app}}$ throughout and do not interpret it as the physical Li concentration at the original surface.

All Hall profiles reported in this work were measured at $77~\mathrm{K}$ to suppress intrinsic carrier contributions in Ge and to obtain active donor concentrations more directly relevant to cryogenic HPGe detector operation~\cite{knoll2010radiation,gilmore2008practical,lutz1999semiconductor,iwan1982highpurity}. No Hall-factor correction or multilayer Hall inversion was applied. Therefore, the reconstructed values are reported as difference-derived apparent Hall donor concentrations. One coupon was measured for each temperature--cooling condition ($n=1$ per condition); the fitted-parameter errors characterize individual profile fits and do not establish coupon-to-coupon or process reproducibility. Systematic effects associated with nonuniform conductivity weighting, residual bulk conduction, contact geometry, visually controlled rather than quantitatively measured Li source loading, and coupon-to-coupon variation remain limitations of the method.

\section{Hall-derived active Li profiles and apparent diffusion coefficients}

\subsection{Effect of cooling protocol: example at 280 $^{\circ}$C}

\begin{figure}[htp!]
    \centering
    \includegraphics[width=1.0\textwidth]{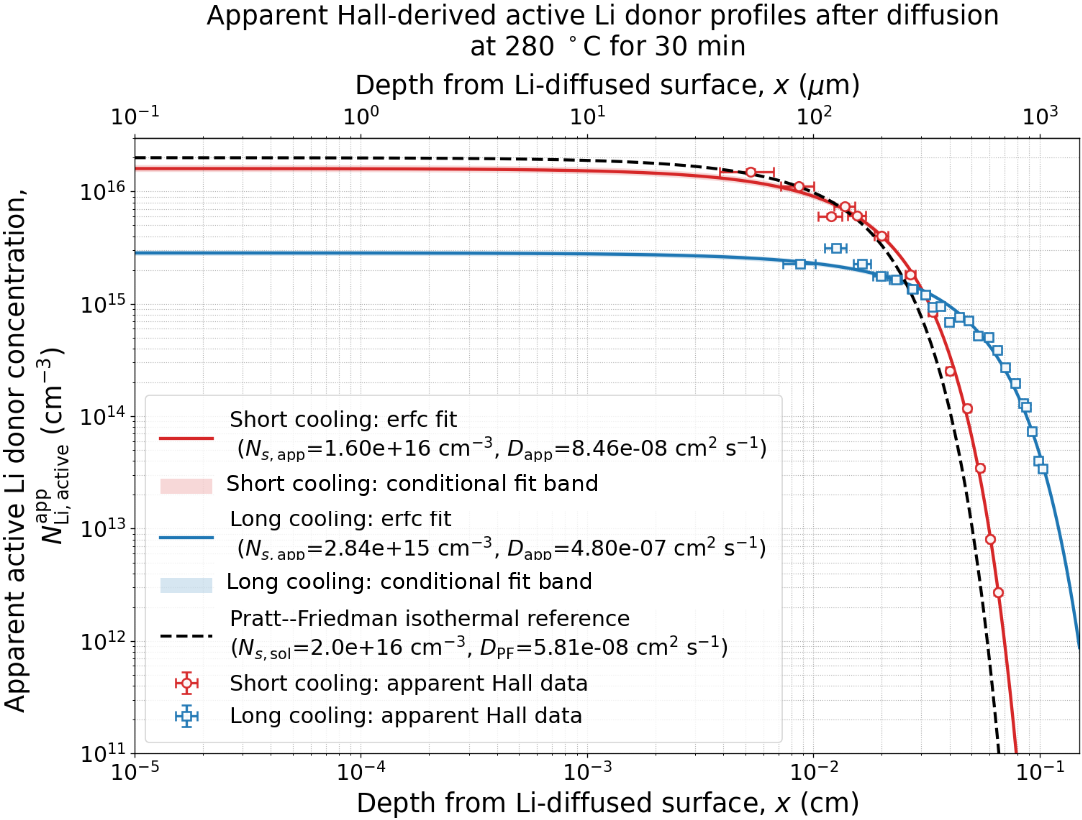}
    \caption{Comparison of difference-derived apparent Hall donor profiles measured at $77~\mathrm{K}$ after Li diffusion at $280~^{\circ}\mathrm{C}$ for $30~\mathrm{min}$ under short- and long-cooling protocols. Data points are assigned to the midpoint depths of finite removed intervals; they are not direct arithmetic slice averages when mobility varies with depth. Solid curves show log-space complementary-error-function fits, and shaded bands show conditional fit-parameter bands under the adopted $5.0\%$ weighting model. The dashed curve is the Pratt--Friedman isothermal reference profile for $280~^{\circ}\mathrm{C}$~\cite{pratt1966diffusion}.}
    \label{fig:cooling_comparison_280C}
\end{figure}

Figure~\ref{fig:cooling_comparison_280C} compares two representative apparent Hall-derived active Li donor profiles measured after Li diffusion at $280~^{\circ}\mathrm{C}$ for $30~\mathrm{min}$. The two coupons were processed with the same nominal diffusion temperature and hold time, but with different post-diffusion cooling protocols. The corresponding measured cooling histories are shown in Fig.~\ref{fig:cooling_history}. With one coupon per condition, Fig.~\ref{fig:cooling_comparison_280C} provides a direct comparison of the two post-diffusion thermal histories but does not establish process reproducibility.

The horizontal error bars in Fig.~\ref{fig:cooling_comparison_280C} represent the propagated depth uncertainty from the digital-caliper thickness measurements. Since each removed depth was obtained from a difference between two averaged thickness measurements, and the caliper-resolution term was not reduced by averaging, the depth uncertainty was conservatively estimated as
\[
    \sigma_x=\sqrt{2}\times0.01~\mathrm{mm}
    =14.1~\mu\mathrm{m}.
\]
For the shallowest points plotted on the logarithmic depth axis, only the displayed lower horizontal error bar was clipped to remain at positive depth; the stored uncertainty value was not changed. The vertical error bars use the $5.0\%$ relative concentration scale adopted in the source analysis workbooks for plotting and log-space fitting weights. This value is a common weighting scale, not a formal propagation of the uncertainties in the two Hall measurements, two thickness measurements, and etch correction entering Eq.~\eqref{eq:hall_adjacent_difference}. In particular, subtraction of similar sheet-density terms can amplify the relative uncertainty of a reconstructed point. The plotted scale therefore should not be interpreted as a complete statistical or systematic uncertainty envelope. Single-measurement Hall repeatability, Hall-factor assumptions, mobility weighting, temperature-profile mismatch between the witness sample and the Li-coated coupon, the distribution and calibration uncertainty of the etch correction, correlations between adjacent reconstructed points, and coupon-to-coupon variation are not included because the required inputs were not available in the archived data and fitting notebooks.

The shaded bands in Fig.~\ref{fig:cooling_comparison_280C} are conditional fit-parameter bands obtained by sampling the covariance matrix of the log-space least-squares fit and recalculating the corresponding erfc profiles. The lower and upper bounds are the 16th and 84th percentiles of the sampled fitted curves. Because the fit uses the adopted $5.0\%$ weighting scale and does not propagate the full adjacent-difference measurement model, these bands quantify fit precision under the stated assumptions rather than total profile uncertainty or process reproducibility.

The two profiles differ substantially even though the nominal diffusion hold was identical. The short-cooling sample retains a high near-surface apparent active donor concentration, with a fitted extrapolated apparent intercept of approximately $1.6\times10^{16}~\mathrm{cm^{-3}}$. In contrast, the long-cooling sample has a lower fitted extrapolated apparent intercept of approximately $2.8\times10^{15}~\mathrm{cm^{-3}}$, while its active donor profile extends to larger depths. For the coupons studied, this behavior shows a strong association between the final Hall-derived active donor profile and the post-hold cooling history, in addition to the nominal isothermal diffusion hold~\cite{fuller1953diffusion,fuller1954mobility,pratt1966diffusion,pell1957solubility,morin1957precipitation,carter1960kinetics}.

For a compact process-level description of the measured profiles, the apparent active donor data were fitted using
\begin{equation}
N_{\mathrm{Li,active}}^{\mathrm{app}}(x)
=
N_{s,\mathrm{app}}
\operatorname{erfc}
\left(
\frac{x}{2\sqrt{D_{\mathrm{app}}t_{\mathrm{hold}}}}
\right),
\label{eq:apparent_erfc_280C}
\end{equation}
where $t_{\mathrm{hold}}=1800~\mathrm{s}$ is the nominal diffusion hold time. In this analysis, $N_{s,\mathrm{app}}$ and $D_{\mathrm{app}}$ are apparent fit parameters for the final Hall-derived active donor profile. They should not be interpreted as the equilibrium Li solubility or the intrinsic isothermal Li diffusivity in Ge~\cite{pell1957solubility,fuller1953diffusion,pratt1966diffusion}. The fits were performed in log-space so that the near-surface and tail regions of the profile both contributed to the fitted parameters.

The short-cooling profile is described by
\[
N_{s,\mathrm{app}}=1.6004\times10^{16}~\mathrm{cm^{-3}},
\qquad
D_{\mathrm{app}}=8.4578\times10^{-8}~\mathrm{cm^{2}\,s^{-1}}.
\]
The long-cooling profile is described by
\[
N_{s,\mathrm{app}}=2.8395\times10^{15}~\mathrm{cm^{-3}},
\qquad
D_{\mathrm{app}}=4.7985\times10^{-7}~\mathrm{cm^{2}\,s^{-1}}.
\]
Because the same nominal hold time is used in Eq.~\eqref{eq:apparent_erfc_280C}, the fitted $D_{\mathrm{app}}$ absorbs the effect of additional thermal exposure during cooling. Therefore, the larger $D_{\mathrm{app}}$ obtained for the long-cooling sample should be understood as an apparent process-level parameter associated with a broader final active donor distribution, not as a direct change in the intrinsic Li diffusivity during the $280~^{\circ}\mathrm{C}$ hold.

The Pratt--Friedman curve in Fig.~\ref{fig:cooling_comparison_280C} is included only as an isothermal literature reference for classical Li diffusion in Ge~\cite{pratt1966diffusion}. In this reference calculation,
\[
N_{s,\mathrm{sol}}=2.0\times10^{16}~\mathrm{cm^{-3}},
\qquad
D_{\mathrm{PF}}=5.81\times10^{-8}~\mathrm{cm^{2}\,s^{-1}},
\qquad
t=1800~\mathrm{s}.
\]
The short-cooling profile is closer to this reference profile than the long-cooling profile, especially in the near-surface apparent donor parameter. However, exact agreement is not expected because the present measurements report mobility-weighted Hall-derived donor values at $77~\mathrm{K}$ after the complete processing and cooling sequence, rather than total Li concentration profiles~\cite{fuller1954mobility,sze2007physics,shur1990physics}. For the coupons studied, the strong difference between the short- and long-cooling profiles supports treating cooling history as an explicit recorded processing variable when Hall-derived profiles are used to evaluate Li diffusion for HPGe detector fabrication.

\subsection{Temperature dependence of active Li profiles}

\begin{figure}[htp!]
    \centering
    \includegraphics[width=1.0\linewidth]{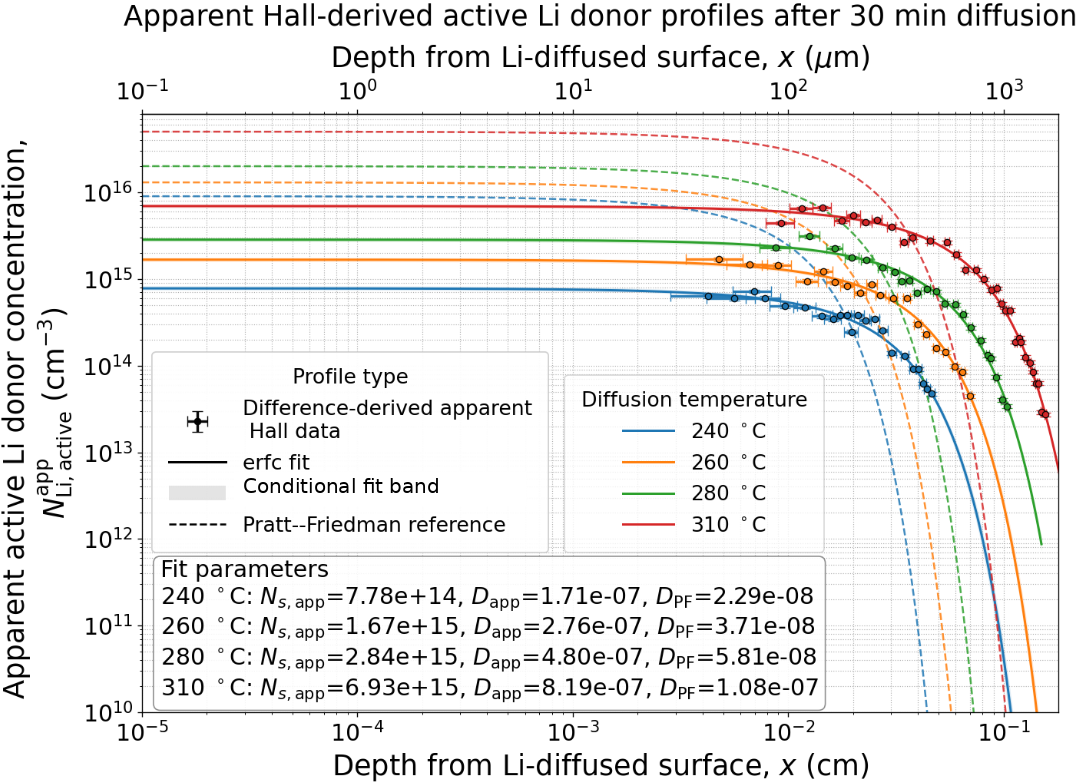}
    \caption{Difference-derived apparent Hall donor profiles measured at $77~\mathrm{K}$ after $30~\mathrm{min}$ Li diffusion under the \emph{long-cooling} protocol. Data points are assigned to finite removed-depth intervals and remain subject to mobility weighting. Solid curves show log-space complementary-error-function fits, and shaded bands show conditional fit-parameter bands under the adopted $5.0\%$ weighting model. Dashed curves show Pratt--Friedman isothermal reference profiles calculated for the corresponding diffusion temperatures~\cite{pratt1966diffusion}. The plotted error bars follow the depth and weighting-scale treatment described in Fig.~\ref{fig:cooling_comparison_280C}.}
    \label{fig:long_cooling_profiles}
\end{figure}

\begin{figure}[htp!]
    \centering
    \includegraphics[width=1.0\linewidth]{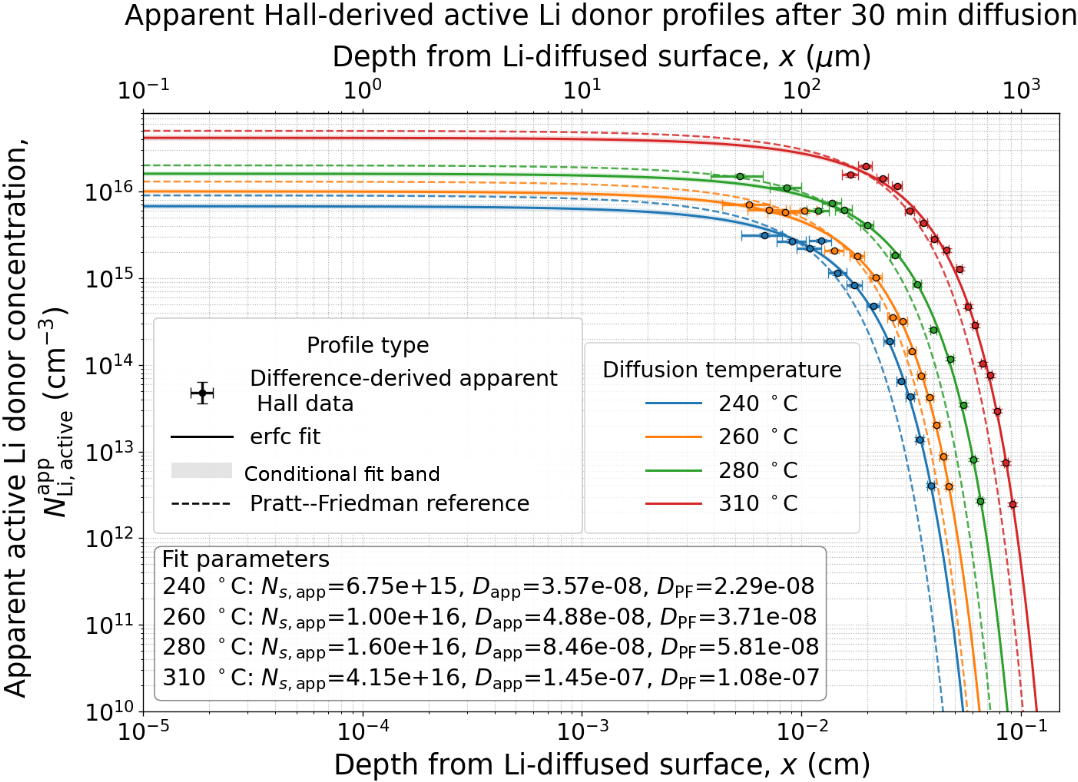}
    \caption{Difference-derived apparent Hall donor profiles measured at $77~\mathrm{K}$ after $30~\mathrm{min}$ Li diffusion under the \emph{short-cooling} protocol. Data points are assigned to finite removed-depth intervals and remain subject to mobility weighting. Solid curves show log-space complementary-error-function fits, and shaded bands show conditional fit-parameter bands under the adopted $5.0\%$ weighting model. Dashed curves show Pratt--Friedman isothermal reference profiles calculated for the corresponding diffusion temperatures~\cite{pratt1966diffusion}. The plotted error bars follow the depth and weighting-scale treatment described in Fig.~\ref{fig:cooling_comparison_280C}.}
    \label{fig:short_cooling_profiles}
\end{figure}

Figures~\ref{fig:long_cooling_profiles} and~\ref{fig:short_cooling_profiles} show the apparent Hall-derived active Li donor profiles measured at $77~\mathrm{K}$ after $30~\mathrm{min}$ Li diffusion at different temperatures. The same data-reduction and fitting procedure described above was applied to both cooling protocols. The plotted points are difference-derived apparent Hall donor concentrations assigned to the corresponding removed-depth intervals; they are mobility-weighted operational quantities rather than direct arithmetic slice averages. For each profile, the data were fitted in log-space using the complementary-error-function form in Eq.~\eqref{eq:apparent_erfc_280C}, with both $N_{s,\mathrm{app}}$ and $D_{\mathrm{app}}$ treated as free apparent fit parameters. The corresponding Pratt--Friedman isothermal reference profile was calculated using the same nominal diffusion temperature and hold time~\cite{pratt1966diffusion}.

For both cooling protocols, increasing the diffusion temperature shifts the fitted active donor profile toward larger depths. This trend is consistent with the expected thermally activated transport of Li in Ge~\cite{fuller1953diffusion,fuller1954mobility,pratt1966diffusion}. Under the long-cooling protocol, the fitted apparent diffusion coefficient increases from $1.7107\times10^{-7}$ to $8.1886\times10^{-7}~\mathrm{cm^{2}\,s^{-1}}$ as the diffusion temperature increases from $240$ to $310~^{\circ}\mathrm{C}$. Over the same temperature range, the extrapolated apparent intercept $N_{s,\mathrm{app}}$ increases from $7.7817\times10^{14}$ to $6.9257\times10^{15}~\mathrm{cm^{-3}}$. Under the short-cooling protocol, $D_{\mathrm{app}}$ increases from $3.5690\times10^{-8}$ to $1.4548\times10^{-7}~\mathrm{cm^{2}\,s^{-1}}$, while $N_{s,\mathrm{app}}$ increases from $6.7477\times10^{15}$ to $4.1502\times10^{16}~\mathrm{cm^{-3}}$.

A clear difference is observed between the two cooling protocols across the full temperature range. At each diffusion temperature, the short-cooling profiles retain substantially larger near-surface apparent active donor concentrations, whereas the long-cooling profiles show lower fitted $N_{s,\mathrm{app}}$ values and broader low-concentration tails. In the fitted parameter sets shown in the figures, the long-cooling $D_{\mathrm{app}}$ values are consistently larger than the corresponding short-cooling values, while the long-cooling $N_{s,\mathrm{app}}$ values are consistently lower. For the coupons studied, this systematic association is consistent with the additional thermal exposure during slow cooling contributing to the final electrically active donor distribution rather than simply preserving the profile formed during the nominal isothermal hold~\cite{fuller1953diffusion,pratt1966diffusion,pell1957solubility,morin1957precipitation,carter1960kinetics}.

The Pratt--Friedman curves in Figs.~\ref{fig:long_cooling_profiles} and~\ref{fig:short_cooling_profiles} are included as classical isothermal references~\cite{pratt1966diffusion}. They are not expected to reproduce the measured profiles exactly because the present data represent Hall-derived electrically active donor concentrations at $77~\mathrm{K}$ after complete processing and cooling histories. In particular, Hall measurements are sensitive to electrically active carriers and do not directly measure the total Li concentration~\cite{fuller1954mobility,sze2007physics,shur1990physics}. Differences between the measured profiles and the isothermal reference curves may therefore arise from the non-isothermal cooling history, changes in the electrically active fraction of Li, Li redistribution, precipitation, compensation, or other process-dependent effects.

The lower $N_{s,\mathrm{app}}$ values obtained after long cooling should not be interpreted as evidence for a single microscopic mechanism. Several mechanisms can produce the observed combination of a reduced near-surface apparent active donor concentration and an extended low-concentration tail. First, because Li remains highly mobile in Ge at elevated temperature, the additional thermal exposure during slow cooling can redistribute Li from the near-surface region into the bulk, thereby broadening the profile while reducing the fitted extrapolated apparent intercept~\cite{fuller1953diffusion,fuller1954mobility,pratt1966diffusion}. Second, as the sample cools through a changing solubility condition, part of the Li population may precipitate or become incorporated into electrically inactive Li-related complexes. Previous studies of Li in Ge have shown that Li solubility, precipitation, ion-pair or ion-triplet formation, and phase-equilibrium behavior can depend strongly on thermal history, impurity content, and crystal condition~\cite{pell1957solubility,morin1957precipitation,reiss1958effect,carter1960kinetics,sangster1997geli}. Third, compensation by residual acceptors or thermally generated defects could reduce the net electrically active donor concentration measured by Hall effect without requiring an equivalent decrease in the total Li concentration. Additional possibilities include Li out-diffusion, surface reactions during the longer high-temperature exposure, condition-to-condition variation in the visually applied Li source loading, or artifacts associated with the Hall-depth-profiling method.

These statements are interpretations of the observed Hall-derived profile differences rather than direct microscopic identifications. The experimental observation is the difference between the short- and long-cooling Hall-derived profiles; the erfc fit provides an empirical parameterization of that difference; redistribution, deactivation, compensation, source-loading variation, and related processes are possible contributing mechanisms. The present Hall-derived profiles cannot distinguish among these possibilities. The Hall measurement is sensitive to electrically active carriers under the measurement conditions and, for a nonuniform diffused layer, the reconstructed profile remains a difference-derived, mobility-weighted quantity rather than a direct local or total-Li concentration profile~\cite{schroder2006semiconductor,basol2024dhem}. Therefore, the long-cooling result should be described as a reduction in the extrapolated \emph{Hall-derived active donor parameter}, not as direct proof of Li loss from the near-surface region. A more complete physical separation of the mechanisms would require complementary measurements. For example, secondary-ion mass spectrometry combined with electrical profiling could separate the total Li distribution from the electrically active carrier distribution~\cite{bennett2016simsar}, while capacitance--voltage profiling, spreading-resistance profiling, differential Hall profiling, and detector charge-collection or dead-layer measurements could test the electrical and detector-response consequences of the modified Li-diffused region~\cite{schroder2006semiconductor,basol2024dhem,aguayo2013signals}.

Across the full data set, the final apparent active donor profile is associated with both the nominal diffusion temperature and the post-diffusion cooling protocol. The erfc form remains useful for summarizing each measured profile~\cite{crank1975mathematics,shewmon1989diffusion,mehrer2007diffusion}, but the extracted $D_{\mathrm{app}}$ and $N_{s,\mathrm{app}}$ should be regarded as apparent process-level parameters for the final Hall-derived active donor distribution. The temperature dependence of $D_{\mathrm{app}}$ is analyzed in the next subsection and compared with classical Arrhenius descriptions of Li diffusion in Ge~\cite{fuller1953diffusion,fuller1954mobility,pratt1966diffusion}.

\subsection{Apparent diffusion coefficients from erfc-profile fitting}

To compare the temperature dependence of the measured active Li donor profiles in a compact form, an apparent diffusion coefficient, $D_{\mathrm{app}}$, was extracted from the complementary-error-function representation described in Eq.~\eqref{eq:apparent_erfc_280C}~\cite{crank1975mathematics,shewmon1989diffusion,mehrer2007diffusion}. For each diffusion temperature and cooling protocol, $D_{\mathrm{app}}$ describes the width of the final Hall-derived electrically active donor profile after the complete processing cycle. Therefore, $D_{\mathrm{app}}$ is treated here as an apparent process-level parameter, not as the intrinsic isothermal diffusivity of Li in Ge~\cite{fuller1953diffusion,fuller1954mobility,pratt1966diffusion}.

\begin{figure}[htp!]
    \centering
    \includegraphics[width=1.0\linewidth]{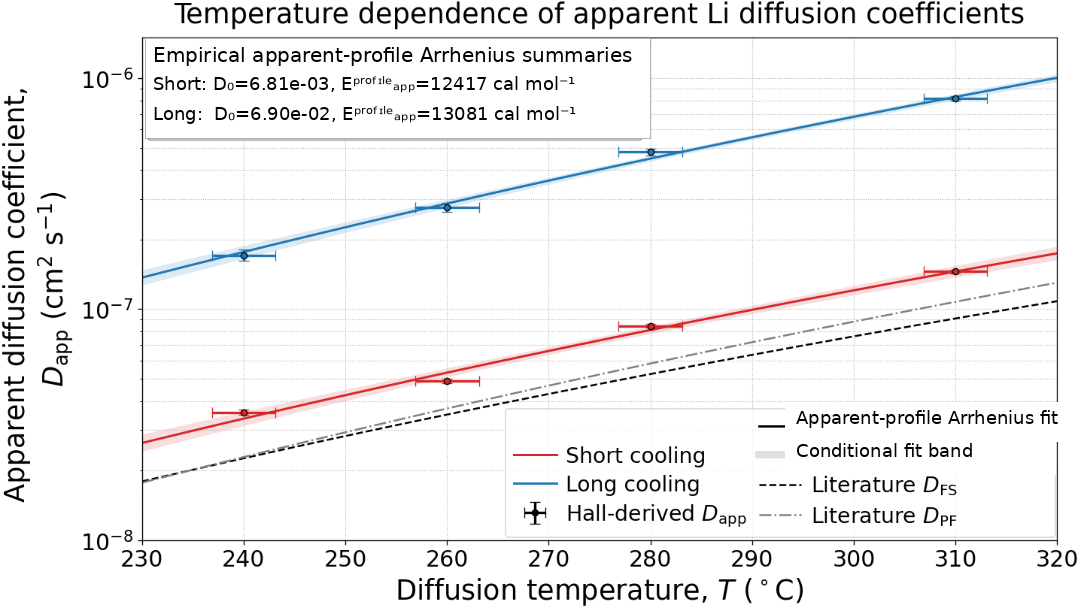}
    \caption{Temperature dependence of the apparent diffusion coefficient $D_{\mathrm{app}}$ extracted from complementary-error-function fits to the Hall-derived active Li donor profiles. Symbols represent the extracted $D_{\mathrm{app}}$ values for the short- and long-cooling protocols. Vertical error bars show conditional asymmetric fit-parameter errors under the adopted weighting model; they are not process-reproducibility intervals. Horizontal error bars show an instrument-specification-based thermometer/probe scale of $\pm 3.13~^{\circ}\mathrm{C}$ around the nominal set points. They do not represent the uncertainty of a retained time-averaged coupon temperature and do not include possible witness-to-coupon temperature offsets. Solid lines show empirical apparent-profile Arrhenius summaries of each cooling data set, and shaded bands show approximate one-standard-deviation parameter bands for those empirical summaries. The dashed and dash-dotted curves show the literature isothermal diffusion relations of Fuller and Severiens and Pratt and Friedman, calculated using Eqs.~\eqref{eq:fs_ge} and~\eqref{eq:pf_ge}, respectively~\cite{fuller1954mobility,pratt1966diffusion}.}
    \label{fig:Dapp_vs_temperature}
\end{figure}

Figure~\ref{fig:Dapp_vs_temperature} summarizes the extracted $D_{\mathrm{app}}$ values for the short- and long-cooling data sets. For both cooling protocols, $D_{\mathrm{app}}$ increases with diffusion temperature, consistent with the expected thermally activated transport of Li in Ge~\cite{fuller1953diffusion,fuller1954mobility,pratt1966diffusion}. Under the short-cooling protocol, $D_{\mathrm{app}}$ increases from $3.569\times10^{-8}$ to $1.455\times10^{-7}~\mathrm{cm^{2}\,s^{-1}}$ as the diffusion temperature increases from $240$ to $310~^{\circ}\mathrm{C}$. Under the long-cooling protocol, the corresponding values are larger, increasing from $1.711\times10^{-7}$ to $8.189\times10^{-7}~\mathrm{cm^{2}\,s^{-1}}$ over the same temperature range.

At every diffusion temperature, the long-cooling protocol gives a larger $D_{\mathrm{app}}$ than the short-cooling protocol. This trend is consistent with the profile comparisons in Figs.~\ref{fig:long_cooling_profiles} and~\ref{fig:short_cooling_profiles}, where the long-cooling samples show broader active donor distributions extending to larger depths. The larger $D_{\mathrm{app}}$ values are consistent with additional thermal exposure during cooling being incorporated into the final measured active profile. They do not imply that the intrinsic Li diffusivity during the nominal isothermal hold is different for the two cooling protocols.

For visual comparison only, the temperature dependence of each apparent-profile data set was fitted with an Arrhenius-type expression~\cite{fuller1953diffusion,fuller1954mobility,pratt1966diffusion},
\begin{equation}
D_{\mathrm{app}}(T)
=
D_{0,\mathrm{app}}
\exp\left(-\frac{E_{\mathrm{app}}^{\mathrm{profile}}}{RT}\right),
\label{eq:arrhenius_dapp}
\end{equation}
where $T$ is the absolute temperature and $R=1.98~\mathrm{cal~mol^{-1}~K^{-1}}$. The empirical Arrhenius summaries are
\begin{equation}
D_{\mathrm{app}}^{\mathrm{short}}(T)
=
6.814\times10^{-3}
\exp\left(-\frac{12417}{RT}\right)
~\mathrm{cm^{2}\,s^{-1}},
\label{eq:arrhenius_short}
\end{equation}
and
\begin{equation}
D_{\mathrm{app}}^{\mathrm{long}}(T)
=
6.901\times10^{-2}
\exp\left(-\frac{13081}{RT}\right)
~\mathrm{cm^{2}\,s^{-1}}.
\label{eq:arrhenius_long}
\end{equation}
The fitted energy-like quantity in Eq.~\eqref{eq:arrhenius_dapp} is therefore denoted $E_{\mathrm{app}}^{\mathrm{profile}}$ and is referred to as an \emph{empirical apparent-profile activation parameter}. It is not a measurement of the microscopic Li migration barrier or intrinsic diffusion activation energy. The fits were performed in logarithmic space. The corresponding log-space goodness-of-fit values were $\mathrm{RMSE}_{\log_{10}}=0.0241$ and $R^2_{\log_{10}}=0.9893$ for the short-cooling data set, and $\mathrm{RMSE}_{\log_{10}}=0.0184$ and $R^2_{\log_{10}}=0.9948$ for the long-cooling data set. Because only four nominal diffusion temperatures enter each empirical summary and $D_{\mathrm{app}}$ includes the complete post-hold processing history, Eqs.~\eqref{eq:arrhenius_short} and~\eqref{eq:arrhenius_long} are not independent determinations of fundamental Li transport parameters.

The extracted short-cooling $D_{\mathrm{app}}$ values are close to, but generally above, the classical isothermal reference curves. The long-cooling $D_{\mathrm{app}}$ values are higher than both literature references across the full temperature range studied. This difference is not treated as a disagreement with the classical isothermal diffusion coefficients. Instead, it reflects the fact that the present $D_{\mathrm{app}}$ values are extracted from final Hall-derived active donor profiles after complete thermal cycles, including post-diffusion cooling. In contrast, the literature curves represent idealized isothermal Li diffusion behavior~\cite{fuller1953diffusion,fuller1954mobility,pratt1966diffusion}.

These results show that the erfc-profile fit is useful for parameterizing the measured active donor profiles, but the extracted apparent diffusion coefficient differs systematically between the two cooling protocols for the coupons studied. The empirical apparent-profile Arrhenius behavior of $D_{\mathrm{app}}$ therefore provides only a process-level summary of the final active Li distribution, while also showing that the measured cooling history should be considered when estimating Li-diffused contact profiles for HPGe detector fabrication.

Table~\ref{tab:fit_summary} lists the sample ID, nominal diffusion set point, endpoint cooling time, and fit results for all eight coupons. Each sample ID combines the nominal set point with the assigned cooling protocol. Each row represents one coupon; therefore, the replicate number is $n=1$ for every condition. The asymmetric parameter errors and goodness-of-fit statistics quantify the conditional erfc fit to that profile under the adopted weighting scale. They are not estimates of process reproducibility.

\begin{table}[htbp]
\centering
\scriptsize
\setlength{\tabcolsep}{3.2pt}
\renewcommand{\arraystretch}{1.12}
\caption{Condition-level summary for the eight measured coupons. $N_{\mathrm{rem}}$ is the number of sequential removal steps and $N_{\mathrm{fit}}$ is the number of valid difference-derived profile points entering the erfc fit. Sample IDs encode the nominal diffusion set point and cooling protocol. $T_{\mathrm{nom}}$ values are nominal process set points, not independently measured time-averaged coupon temperatures. Each condition contains one coupon ($n=1$). Parameter errors are conditional asymmetric fit errors, not coupon-to-coupon reproducibility intervals.}
\label{tab:fit_summary}
\resizebox{\textwidth}{!}{%
\begin{tabular}{@{}l c r r r c c c c@{}}
\hline
Sample ID & $T_{\mathrm{nom}}$ & $t_{\mathrm{cool},30}$ & $N_{\mathrm{rem}}$ & $N_{\mathrm{fit}}$ & $N_{s,\mathrm{app}}$ & $D_{\mathrm{app}}$ & $\mathrm{RMSE}_{\log_{10}}$ & $R^2_{\log_{10}}$ \\
 & $({}^{\circ}\mathrm{C})$ & $(\mathrm{min})$ &  &  & $(\mathrm{cm^{-3}})$ & $(\mathrm{cm^2\,s^{-1}})$ &  &  \\
\hline
240C-Short & 240 & 6.50 & 13 & 12 & $6.748^{+0.561}_{-0.518}\times10^{15}$ & $3.569^{+0.093}_{-0.091}\times10^{-8}$ & 0.0671 & 0.9949 \\
240C-Long  & 240 & 95.67 & 23 & 22 & $7.782^{+0.446}_{-0.422}\times10^{14}$ & $1.711^{+0.097}_{-0.091}\times10^{-7}$ & 0.0672 & 0.9652 \\
260C-Short & 260 & 6.65 & 16 & 15 & $1.000^{+0.073}_{-0.068}\times10^{16}$ & $4.884^{+0.102}_{-0.100}\times10^{-8}$ & 0.0690 & 0.9958 \\
260C-Long  & 260 & 97.43 & 20 & 19 & $1.669^{+0.097}_{-0.092}\times10^{15}$ & $2.761^{+0.128}_{-0.122}\times10^{-7}$ & 0.0672 & 0.9786 \\
280C-Short & 280 & 6.80 & 14 & 13 & $1.600^{+0.116}_{-0.108}\times10^{16}$ & $8.458^{+0.170}_{-0.167}\times10^{-8}$ & 0.0696 & 0.9967 \\
280C-Long  & 280 & 99.55 & 23 & 22 & $2.840^{+0.160}_{-0.152}\times10^{15}$ & $4.799^{+0.162}_{-0.157}\times10^{-7}$ & 0.0673 & 0.9855 \\
310C-Short & 310 & 7.17 & 17 & 16 & $4.150^{+0.328}_{-0.304}\times10^{16}$ & $1.455^{+0.027}_{-0.027}\times10^{-7}$ & 0.0739 & 0.9962 \\
310C-Long  & 310 & 103.15 & 32 & 31 & $6.926^{+0.321}_{-0.307}\times10^{15}$ & $8.189^{+0.169}_{-0.166}\times10^{-7}$ & 0.0673 & 0.9917 \\
\hline
\end{tabular}%
}
\end{table}

\subsection{Fit-derived concentration-threshold depths}
\label{sec:detector_relevant_threshold_depths}

For HPGe detector fabrication, the fitted active Li donor profiles are also useful when reduced to depth metrics that can be compared with detector-contact and active-volume length scales. Although an apparent Hall-derived donor profile does not by itself define an abrupt dead-layer boundary, the depth at which the fitted profile falls below selected active-donor concentrations provides a practical process-level metric for comparing Li-diffused contacts. In particular, threshold depths near $10^{14}$--$10^{12}~\mathrm{cm^{-3}}$ are relevant to estimating the extent of the highly doped contact, the transition region between the $n^{+}$ contact and the HPGe bulk, and the depth range that may require special treatment in detector-response simulations~\cite{jiang2016deadlayer,ma2017inactive,dai2023modeling}.

For each fitted profile, the threshold depth was calculated from the erfc fit as
\begin{equation}
x(N_{\mathrm{th}})
=
2\sqrt{D_{\mathrm{app}}t_{\mathrm{hold}}}
\,
\mathrm{erfc}^{-1}
\left(
\frac{N_{\mathrm{th}}}{N_{s,\mathrm{app}}}
\right),
\label{eq:threshold_depth}
\end{equation}
where $N_{\mathrm{th}}$ is the selected apparent active donor concentration threshold and $t_{\mathrm{hold}}=1800~\mathrm{s}$. The resulting depths are listed in Table~\ref{tab:threshold_depths}. These values are derived from the fitted apparent profiles and should therefore be interpreted as process-comparison metrics, not as direct measurements of abrupt electrical or charge-collection boundaries. No confidence intervals are assigned because the present weighting-scale analysis does not propagate the correlated Hall, thickness, and etch-correction uncertainties through the adjacent-difference reconstruction.

\begin{table}[t]
\centering
\small
\setlength{\tabcolsep}{7pt}
\renewcommand{\arraystretch}{1.05}
\caption{Fit-derived concentration-threshold depths calculated from the apparent active Li profiles. Here $x_{14}$, $x_{13}$, and $x_{12}$ denote the depths at which the fitted profile falls below $10^{14}$, $10^{13}$, and $10^{12}~\mathrm{cm^{-3}}$, respectively. These values are process-comparison metrics, not measurements of detector dead-layer or charge-collection boundaries. The asterisk marks a fit extrapolation that exceeds the nominal $1.8~\mathrm{mm}$ starting coupon thickness.}
\label{tab:threshold_depths}
\begin{tabular}{@{}c l r r r@{}}
\hline
$T$ & Protocol & $x_{14}$ & $x_{13}$ & $x_{12}$ \\
$({}^{\circ}\mathrm{C})$ &  & \multicolumn{3}{c}{$(\mu\mathrm{m})$} \\
\hline
240 & Short cooling & 276.2 & 360.3 & 430.1 \\
240 & Long cooling  & 377.2 & 617.4 & 798.9 \\
260 & Short cooling & 341.6 & 436.3 & 515.9 \\
260 & Long cooling  & 593.1 & 866.4 & 1082.0 \\
280 & Short cooling & 477.1 & 596.9 & 698.5 \\
280 & Long cooling  & 875.3 & 1212.8 & 1485.3 \\
310 & Short cooling & 694.4 & 840.3 & 966.5 \\
310 & Long cooling  & 1328.1 & 1729.7 & $2063.5^{*}$ \\
\hline
\end{tabular}
\end{table}

The threshold-depth comparison shows that, for the present fits, the long-cooling profile extends deeper than the corresponding short-cooling profile at all four diffusion temperatures and at all three selected concentration thresholds. The difference becomes especially pronounced toward the low-concentration tail. This behavior is consistent with the fitted profiles in Figs.~\ref{fig:long_cooling_profiles} and~\ref{fig:short_cooling_profiles}: the long-cooling fit has a lower extrapolated apparent intercept but a substantially broader profile. Because the threshold depths are calculated from the erfc parameterization, values outside the experimentally sampled depth range are extrapolations; in particular, the $x_{12}=2063.5~\mu\mathrm{m}$ value for the $310~^{\circ}\mathrm{C}$ long-cooling fit exceeds the nominal $1.8~\mathrm{mm}$ starting coupon thickness and should not be interpreted as a directly measured depth.

These threshold depths provide fitted profile length scales that can inform detector modeling, but they are not themselves measurements of a detector inactive or dead layer. In detector simulations, the Li-diffused contact is often represented by a near-surface region with reduced or incomplete charge collection, followed by a transition into the active HPGe bulk~\cite{jiang2016deadlayer,ma2017inactive,dai2023modeling}. The values in Table~\ref{tab:threshold_depths} provide fit-derived length-scale estimates for that transition region under the processing conditions studied here, subject to the extrapolation limitation noted above. For the coupons studied, they also show an association between the post-diffusion cooling condition, the extrapolated apparent intercept, and the depth over which the fitted donor tail extends into the detector bulk.

\subsection{Fit-derived apparent active-donor dose}
\label{sec:apparent_dose}

The complementary-error-function parameterization also provides an integrated apparent active-donor dose,
\begin{equation}
Q_{\mathrm{app}}
=
\int_0^{\infty}
N_{s,\mathrm{app}}
\operatorname{erfc}
\left(
\frac{x}{2\sqrt{D_{\mathrm{app}}t_{\mathrm{hold}}}}
\right)
\,dx
=
2N_{s,\mathrm{app}}
\sqrt{\frac{D_{\mathrm{app}}t_{\mathrm{hold}}}{\pi}}.
\label{eq:apparent_dose}
\end{equation}
Because it is calculated from the mobility-weighted apparent erfc fit, $Q_{\mathrm{app}}$ is not a chemical Li dose. It is also more model dependent than the individual reconstructed profile points: the integral includes the extrapolated $x=0$ intercept and extends mathematically beyond the experimentally sampled depth range. It is used here only to test whether the \emph{fitted apparent profiles} can be described as profile broadening at constant Hall-derived apparent active-donor dose.

\begin{table}[htbp]
\centering
\small
\setlength{\tabcolsep}{8pt}
\caption{Model-dependent, fit-derived apparent active-donor dose from Eq.~\eqref{eq:apparent_dose}. The integral is evaluated from the erfc parameterization rather than directly from a chemically measured Li inventory. The ratio compares the long- and short-cooling fits at the same nominal diffusion temperature.}
\label{tab:apparent_dose}
\begin{tabular}{@{}c r r r@{}}
\hline
$T$ & $Q_{\mathrm{app}}^{\mathrm{short}}$ & $Q_{\mathrm{app}}^{\mathrm{long}}$ & Long/short \\
$({}^{\circ}\mathrm{C})$ & \multicolumn{2}{c}{$(\mathrm{cm^{-2}})$} & $(\%)$ \\
\hline
240 & $6.103\times10^{13}$ & $1.541\times10^{13}$ & 25.3 \\
260 & $1.058\times10^{14}$ & $4.199\times10^{13}$ & 39.7 \\
280 & $2.228\times10^{14}$ & $9.416\times10^{13}$ & 42.3 \\
310 & $7.578\times10^{14}$ & $3.000\times10^{14}$ & 39.6 \\
\hline
\end{tabular}
\end{table}

For the present fits, the long-cooling erfc parameterization gives only $25$--$42\%$ of the corresponding short-cooling $Q_{\mathrm{app}}$. Thus, within the erfc representation, the two fitted profiles cannot be related by simple broadening at constant Hall-derived apparent active-donor dose. The decrease may reflect a reduced electrically active fraction through precipitation, complex formation, or compensation; Li out-diffusion or surface loss; coupon-to-coupon variation; or mobility-weighting and adjacent-difference artifacts. The present Hall data cannot separate these mechanisms, and the result should not be interpreted as a measurement of total Li loss.

\section{Descriptive visualization between the measured cooling endpoints}
\label{sec:endpoint_visualization}

The primary results of this work are the measured short- and long-cooling Hall-derived profiles reported in Section~3. Only two cooling endpoints were measured at each nominal diffusion temperature. Therefore, the continuous curves presented in this section are included solely as an engineering visualization of the parameter difference between those measured endpoints; they do not establish the functional dependence of the profile on cooling time.

For a fixed nominal diffusion temperature $T_{\mathrm{hold}}$, define
\begin{equation}
u=
\frac{t_{\mathrm{cool},30}-t_{\mathrm{cool},30}^{\mathrm{short}}}
{t_{\mathrm{cool},30}^{\mathrm{long}}-t_{\mathrm{cool},30}^{\mathrm{short}}},
\qquad 0\le u\le 1.
\label{eq:cooling_progress}
\end{equation}
For visualization only, $D_{\mathrm{app}}$ is linearly interpolated between the two measured fitted values,
\begin{equation}
D_{\mathrm{app}}(u)
=
(1-u)D_{\mathrm{app}}^{\mathrm{short}}
+
uD_{\mathrm{app}}^{\mathrm{long}},
\label{eq:Dapp_prediction_model}
\end{equation}
while the extrapolated apparent intercept is interpolated in logarithmic concentration space,
\begin{equation}
\ln N_{s,\mathrm{app}}(u)
=
(1-u)\ln N_{s,\mathrm{app}}^{\mathrm{short}}
+
u\ln N_{s,\mathrm{app}}^{\mathrm{long}}.
\label{eq:Nsapp_interpolation_model}
\end{equation}
The corresponding descriptive apparent profile is
\begin{equation}
N_{\mathrm{Li,active}}^{\mathrm{interp}}(x,u)
=
N_{s,\mathrm{app}}(u)
\operatorname{erfc}
\left[
\frac{x}{2\sqrt{D_{\mathrm{app}}(u)t_{\mathrm{hold}}}}
\right].
\label{eq:predicted_active_profile}
\end{equation}
Equations~\eqref{eq:cooling_progress}--\eqref{eq:predicted_active_profile} pass through the two measured fitted endpoints by construction. They are not a transport model, are not validated at intermediate cooling times, and should not be extrapolated to other diffusion temperatures, hold times, or cooling trajectories. In particular, $t_{\mathrm{cool},30}$ remains only a descriptive coordinate; the full measured $T(t)$ history in Fig.~\ref{fig:cooling_history}, rather than $t_{\mathrm{cool},30}$ alone, is the physically relevant thermal record.

\begin{figure}[htp!]
    \centering
    \begin{minipage}[t]{0.49\linewidth}
        \centering
        \includegraphics[width=\linewidth]{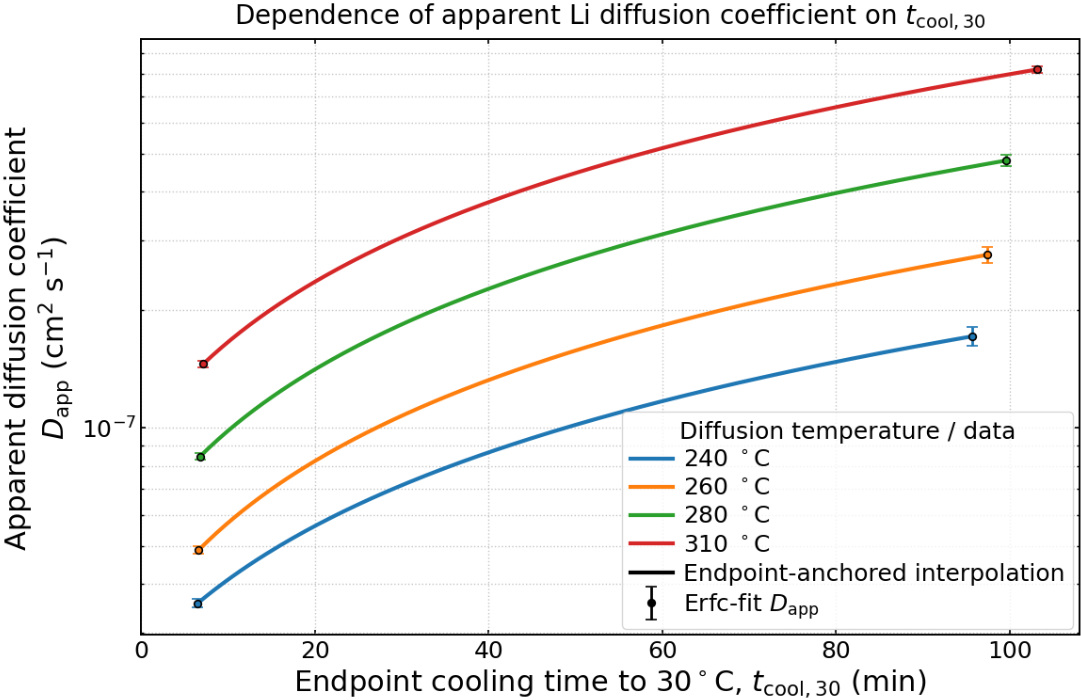}
        \smallskip
        \textbf{(a)}
    \end{minipage}
    \hfill
    \begin{minipage}[t]{0.49\linewidth}
        \centering
        \includegraphics[width=\linewidth]{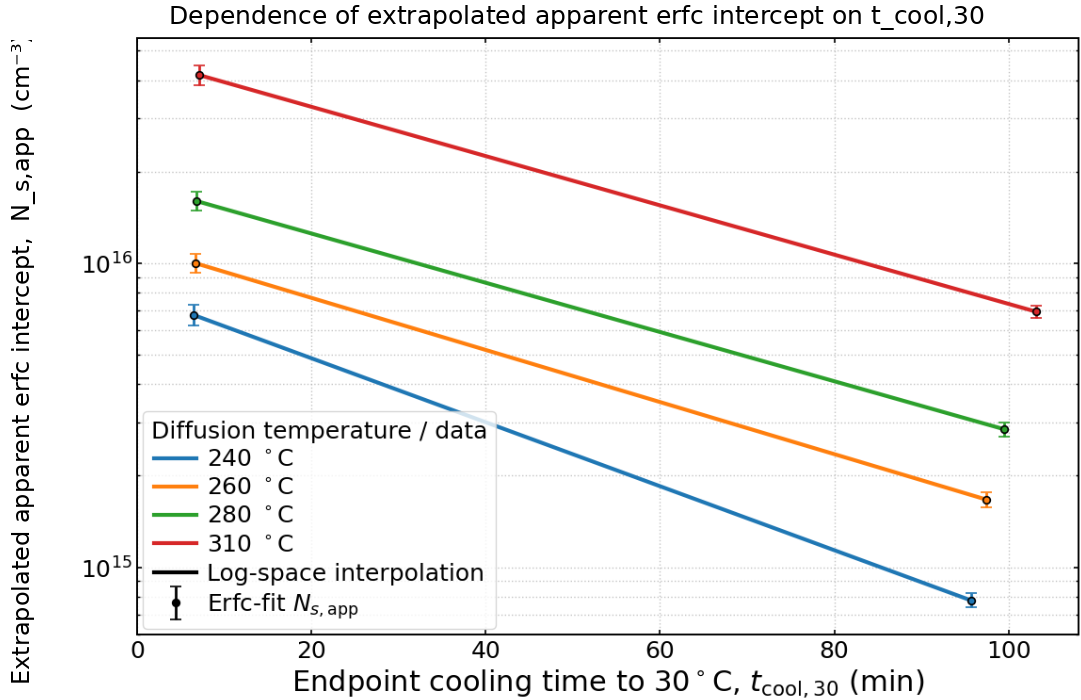}
        \smallskip
        \textbf{(b)}
    \end{minipage}
    \caption{\textbf{Descriptive interpolation only.}
    (a) Measured endpoint values of the apparent profile-width parameter
    $D_{\mathrm{app}}$; solid curves are the endpoint-anchored interpolation of
    Eq.~\eqref{eq:Dapp_prediction_model} and are not independently measured
    cooling-time dependences. Vertical error bars are conditional single-profile
    fit errors, not process-reproducibility intervals.
    (b) Measured endpoint values of the \emph{extrapolated apparent erfc intercept}
    $N_{s,\mathrm{app}}$; solid curves are the log-space endpoint interpolation of
    Eq.~\eqref{eq:Nsapp_interpolation_model} and do not represent measured
    intermediate cooling conditions. Panel (b) identifies $N_{s,\mathrm{app}}$ explicitly as an extrapolated erfc
    intercept rather than a measured surface concentration.}
    \label{fig:endpoint_parameter_visualization}
\end{figure}

At each nominal diffusion temperature, the measured long-cooling endpoint has a larger $D_{\mathrm{app}}$ and a lower $N_{s,\mathrm{app}}$ than the corresponding short-cooling endpoint. The endpoint ratios are approximately $4.8$--$5.7$ for $D_{\mathrm{app}}$, while $N_{s,\mathrm{app}}$ is lower by factors of approximately $5.6$--$8.7$. These statements refer to the measured fitted endpoints; the connecting curves add no independent experimental information.

\begin{figure}[htp!]
    \centering
    \includegraphics[width=1\linewidth]{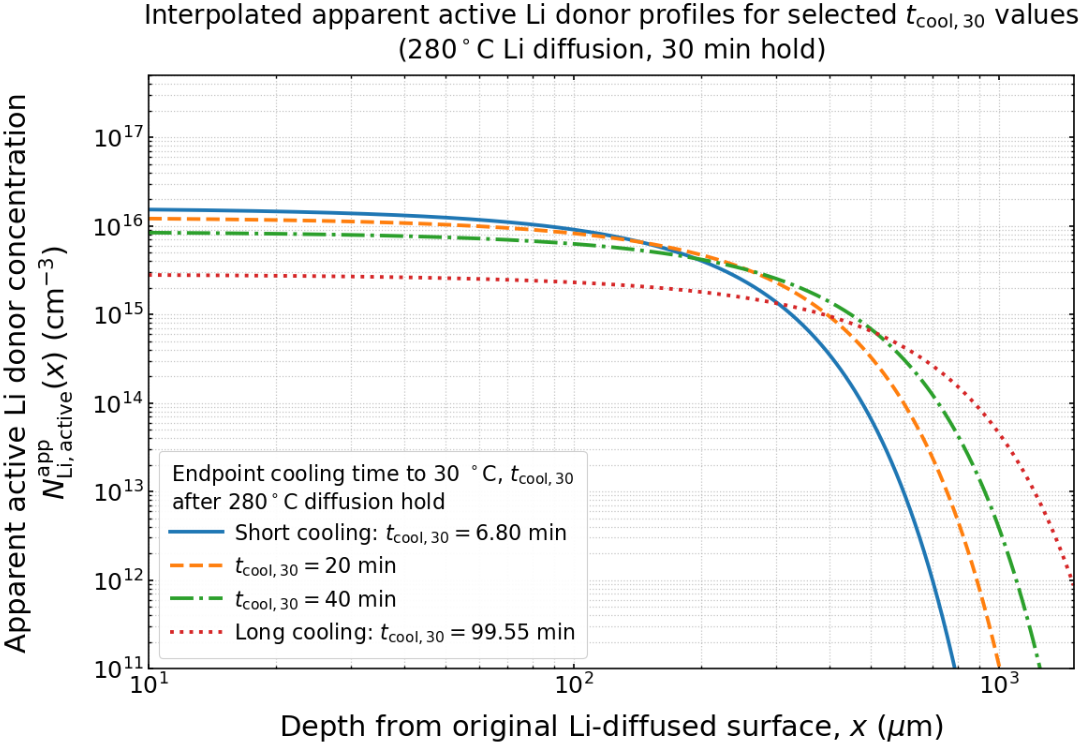}
    \caption{\textbf{Descriptive interpolation only.} Apparent active Li donor profiles for the $280~^{\circ}\mathrm{C}$, $30~\mathrm{min}$ hold generated from Eqs.~\eqref{eq:Dapp_prediction_model}--\eqref{eq:predicted_active_profile}. The $6.80$ and $99.55~\mathrm{min}$ curves reproduce the measured short- and long-cooling fitted endpoints. The $20$ and $40~\mathrm{min}$ curves are hypothetical visualizations and are not independently measured Hall profiles. Curve crossings arise from the chosen erfc parameterization as $N_{s,\mathrm{app}}$ decreases while $D_{\mathrm{app}}$ increases; they are not measured physical boundaries.}
    \label{fig:predicted_profiles_280C}
\end{figure}

Figure~\ref{fig:predicted_profiles_280C} illustrates how the two chosen interpolation rules combine for the $280~^{\circ}\mathrm{C}$ case. The short- and long-cooling curves correspond to the measured fitted endpoints at $t_{\mathrm{cool},30}=6.80$ and $99.55~\mathrm{min}$, whereas the $20$ and $40~\mathrm{min}$ curves are hypothetical interpolation examples only.

The combined endpoint visualization compactly illustrates the opposite fitted-parameter trends. All scientific interpretation in this work is based on the measured endpoint profiles rather than on the interpolation itself.

\section{Summary}

In this work, Li diffusion in HPGe coupon samples was studied using a lithium-in-oil application method~\cite{dong2026hybrid}, a fixed nominal diffusion hold of $30~\mathrm{min}$, nominal temperatures from $240$ to $310~^{\circ}\mathrm{C}$, and two measured post-diffusion cooling protocols. Hall-effect measurements at $77~\mathrm{K}$ combined with sequential material removal were used to estimate difference-derived apparent Hall-active Li donor profiles as a function of depth~\cite{sze2007physics,shur1990physics,lutz1999semiconductor}. Because the diffused layers are nonuniform and mobility weighted in Hall response, the reconstructed profiles are operational electrically active-donor metrics rather than direct local or total-Li concentration profiles.

For the coupons studied, the short- and long-cooling conditions show a consistent association with different final Hall-derived profiles across all four nominal diffusion temperatures. Short cooling is associated with a larger extrapolated erfc intercept $N_{s,\mathrm{app}}$ and a sharper profile, whereas long cooling is associated with a lower $N_{s,\mathrm{app}}$ and a broader low-concentration tail extending farther into the Ge bulk. One coupon was measured for each temperature--cooling condition, so these observations do not quantify run-to-run process reproducibility. The visually controlled Li application, rather than an independently quantified areal source dose for each coupon, is an additional source of possible condition-to-condition variation.

Each measured profile was parameterized with a complementary-error-function form~\cite{crank1975mathematics,shewmon1989diffusion,mehrer2007diffusion}. The resulting $N_{s,\mathrm{app}}$ is an extrapolated fit intercept at $x=0$, not a directly measured surface concentration, and $D_{\mathrm{app}}$ is an apparent final-profile width parameter, not an intrinsic isothermal Li diffusivity. The empirical Arrhenius fits to $D_{\mathrm{app}}$ are therefore used only as apparent-profile summaries; the associated $E_{\mathrm{app}}^{\mathrm{profile}}$ values are not interpreted as microscopic Li migration barriers.

The measured cooling curves are the relevant thermal records. Because Li transport is thermally activated, the endpoint time $t_{\mathrm{cool},30}$ is not an equivalent diffusion time. Equations~\eqref{eq:cooling_thermal_exposure} and~\eqref{eq:total_thermal_exposure} provide a physically motivated way to express diffusion-weighted thermal exposure from the full $T(t)$ trajectory. In this work they are not used to infer an intrinsic diffusion length because the temperature was recorded on a witness Ge piece and unresolved rapid-transfer transients and witness-to-coupon offsets are not independently quantified. The full cooling trajectories are therefore retained as measured process information, while $t_{\mathrm{cool},30}$ is used only as a descriptive endpoint coordinate.

The uncertainty treatment propagates the digital-caliper resolution into the horizontal depth error bars and uses a $5.0\%$ Hall-derived scale for plotting and log-space weighting. The displayed bands and parameter errors are conditional fit-precision measures under that weighting model; they do not represent a complete statistical or systematic uncertainty envelope or process-reproducibility interval. Mobility weighting, Hall-factor assumptions, adjacent-difference correlations, etch uncertainty, source-loading variation, possible witness-to-coupon temperature differences, and coupon-to-coupon variation remain limitations.

Fit-derived concentration-threshold depths at $10^{14}$, $10^{13}$, and $10^{12}~\mathrm{cm^{-3}}$ provide process-comparison length scales, not direct dead-layer or charge-collection boundaries. The $310~^{\circ}\mathrm{C}$ long-cooling value $x_{12}=2063.5~\mu\mathrm{m}$ exceeds the nominal starting coupon thickness and is explicitly identified as an extrapolation. Likewise, the integrated $Q_{\mathrm{app}}$ is a model-dependent apparent active-donor dose obtained from the erfc fit, including extrapolated regions. The long-cooling fits yield $25$--$42\%$ of the corresponding short-cooling $Q_{\mathrm{app}}$, showing that the two fitted apparent profiles cannot be described simply as broadening at constant Hall-derived apparent dose. This should not be interpreted as a measurement of total Li loss.

The observed profile differences, the erfc parameterization, and the microscopic interpretation are intentionally kept separate. The Hall-derived profiles alone cannot determine whether the lower fitted $N_{s,\mathrm{app}}$ after long cooling results from redistribution, precipitation, compensation, electrically inactive Li-related complexes, Li out-diffusion, surface reactions, source-loading variation, or Hall-depth-profiling artifacts. The principal process-level conclusion is therefore that the nominal diffusion temperature and hold time alone are insufficient to characterize the final Hall-active Li contact profile: the measured post-diffusion thermal history must also be considered in HPGe contact engineering and detector-response modeling.

\section*{CRediT authorship contribution statement}

Kunming Dong: Conceptualization, Methodology, Investigation, Formal analysis, Data curation, Software, Visualization, Writing -- original draft.

Dongming Mei: Supervision, Funding acquisition, Resources, Project administration, Writing -- review and editing.

Anupama Karki: Investigation, Formal analysis, Data curation, Writing -- review and editing.

Patrick Burns: Investigation, Formal analysis, Data curation, Writing -- review and editing.

Sanjay Bhattarai: Methodology, Investigation, Resources, Writing -- review and editing.

\section*{Declaration of competing interest}

The authors declare that they have no known competing financial interests or personal relationships that could have appeared to influence the work reported in this paper.

\section*{Funding}

This work was supported in part by the National Science Foundation under Grant Nos. OISE-1743790 and PHY-2310027, by the U.S. Department of Energy under Grant Nos. DE-SC0024519 and DE-SC0004768, and by a research center supported by the State of South Dakota.

\section*{Data availability}

The processed Hall-profile data, cooling-history data underlying the reported figures, and analysis scripts will be made available by the corresponding author on reasonable request.

\bibliographystyle{elsarticle-num} 
\bibliography{cas-refs}
\end{document}